\documentclass[twocolumn,prb]{revtex4}
\usepackage{amsfonts}
\usepackage{amssymb}
\usepackage{makeidx}
\usepackage{amsmath}
\usepackage{graphicx}
\usepackage{dcolumn}
\usepackage{epstopdf}
\usepackage{bm}

\begin{document}

\title{Microscopic quantum description of second-order nonlinearities in 2D
hexagonal nanostructures beyond the Dirac cone approximation }
\author{H.K. Avetissian$^{1}$, G.F. Mkrtchian$^{1}$, K.G. Batrakov$^{2}$,
S.A. Maksimenko$^{2}$}
\affiliation{$1$ Centre of Strong Fields Physics, Yerevan State University, 0025 Yerevan,
Armenia,\\
$2$ Institute for Nuclear Problems, Belorusian State University, 220050
Minsk, Belarus}

\begin{abstract}
Single layers of hexagonal two-dimensional nanostructures such as graphene,
silicene, and germanene exhibit large carrier Fermi velocities and,
consequently, large light-matter coupling strength making these materials
promising elements for nano-opto-electronics. Although these materials are
centrosymmetric, the spatial dispersion turns out to be quite large allowing
the second-order nonlinear response of such materials to be comparable to
the non-centrosymmetric 2D ones. The second-order response of massless Dirac
fermions has been extensively studied, however a general approach correct
over the full Brillouin zone is lacking so far. To complete this gap, in the
current paper we develop a general quantum-mechanical theory of the in-plane
second-order nonlinear response beyond the Dirac cone approximation and
applicable to the full Brillouin zone of the hexagonal tight-binding
nanostructures. We present explicit calculation of the nonlinear
susceptibility tensor of 2D hexagonal nanostructures applicable to arbitrary
three-wave mixing processes.
\end{abstract}

\date{\today }
\maketitle

\section{ Introduction}

In the last decade, graphene\cite{Nov1,Nov2} and its analogs silicene,\cite%
{Sil1,Sil2,Sil3} germanene,\cite{Germ1,Germ2} and stanene\cite{Stan} have
attracted enormous interest due to their unique electronic and optical
properties. These 2D nanostructures consist of honeycomb lattices of atoms
with sublattices made of A and B sites. Hence, in their original structure,
free-standing honeycomb lattices are centrosymmetric, and even-order
nonlinear effects at such nanostructures--light/wave interaction vanish
within the dipole approximation. The latter is fully justified for the
perpendicular incidence of a pump wave to the nanostructure plane. The
symmetry-allowed odd-order nonlinear optical effects are very strong in
graphene-like nanostructures. For graphene, this is confirmed by the
experimental\cite{3rde1,3rde2,3rde3} and theoretical\cite%
{3rd1,3rd2,3rd3,3rd4,3rd5,3rd6,3rd7,3rd8} investigations of the third
harmonic generation process. These nanomaterials can also serve as an active
medium for the extreme nonlinear optical effects, such as high harmonics
generation.\cite{H1,H2,H3,H4,H5,H6,H7,H8,H9,H10,H11}

For even-order nonlinear optical response, one should break the inversion
symmetry in the mentioned nanostructures. In few-layer graphene, the
inversion symmetry can be broken due to interaction between the layers which
results in a second harmonic generation.\cite{Dean1,Dean2} The inversion
symmetry is also broken at the oblique or in-plane propagation of driving
electromagnetic waves. In this case, one should take into account the
spatial dispersion which results in a non-zero in-plane second-order
susceptibility $\chi ^{(2)}$. The second harmonic generation caused by only
intraband transitions in a free-carrier model has been investigated in Refs.
[\onlinecite{Mikhailov,Glazov,Smirnova}]. The difference-frequency
generation and parametric frequency down-conversion with the emphasis on the
nonlinear generation of surface plasmons have been considered in Refs. [%
\onlinecite{Yao, Tokman}]. The experiment [\onlinecite{Constant}] reported
difference-frequency generation of surface plasmons in graphene.
Electron-electron interaction corrections to Feynman diagrams describing
second- and third-order non-linear-response functions have been investigated
in Ref. [\onlinecite{Rostami}]. Valley polarization-induced second harmonic
generation\cite{val1,val2} is also reported.

In Refs. [\onlinecite{Wang,Cheng}] the full quantum-mechanical theory of the
in-plane second-order nonlinear response beyond the electric dipole
approximation has been developed for graphene-like nanostructures
considering the low-energy dynamics in the $\mathrm{K}_{+}$ and $\mathrm{K}%
_{-}$ valleys. In the recent experiment [\onlinecite{Zhang}] the main
theoretical predictions\cite{Wang, Cheng} have been confirmed. In
particular, Fermi-edge resonances at the second harmonic generation in
graphene were reported, and the calculated magnitude of the effective
second-order nonlinear susceptibility\cite{Wang, Cheng} was also close to
the experimental values. In general, the Dirac cone approximation (DCA) is
valid for photon energies much smaller than nearest-neighbor hopping
transfer energy $\hbar \omega <<\gamma _{0}$. In practice, the DCA for
nonlinear optical response is valid up to energies $\gamma _{0}/2$. For
graphene ($\gamma _{0}\simeq 2.8$ eV), this involves the range of
frequencies from THz to the near-infrared. For silicene and germanene $%
\gamma _{0}\simeq 1$ eV and the DCA is violated for mid-infrared
frequencies. Hence, at visible and deep UV frequencies of driving waves for
graphene and even more for silicene, germanene, and stanene one should have
microscopic theory describing nonlinear interaction beyond the DCA and
applicable to the full Brillouin zone (FBZ) of the hexagonal nanostructure
with tight-binding electronic states. Note that spatial dispersion induced
second-order nonlinear response is nonzero for doped system and at
sufficiently high doping $>0.2\ \mathrm{eV}$ one can omit spin-orbit
coupling in silicene, germanene, and stanene considering those as gapless
hexagonal nanostructures with corresponding lattice spacing $a$ and hopping
transfer energy $\gamma _{0}$.

In the present work, we develop the full quantum-mechanical theory of the
in-plane second-order nonlinear response beyond the DCA and applicable to
the FBZ of a hexagonal tight-binding nanostructure. The resulting nonlinear
susceptibility tensor satisfies all symmetry and permutation properties and
can be applied for the arbitrary wave mixing.

The paper is organized as follows. In Sec. II the Hamiltonian within the
tight-binding approximation and the solution of the master equation for the
density matrix are presented. In Sec. III, we calculate the second-order
susceptibility tensor taking into account the spatial dispersion. Then we
examine the susceptibility tensors for second-order harmonic and
difference/sum-frequency generation processes. In particular, we consider
the plasmon generation at the down-conversion. Finally, conclusions are
given in Sec. IV.

\section{The tight-binding Hamiltonian and perturbative solution of the
master equation for the density matrix}

Let a monolayer nanostructure consisting of a honeycomb lattice (see Fig. 1)
interacts with multicolor electromagnetic radiation. We consider the
interaction with obliquely incident waves. A sketch of the interaction
geometry is shown in Fig. 1(c). In the $z$-direction, we have a strong
binding of the electrons. Hence, we will neglect the in-plane component of
the magnetic field or out of the plane electrical field component. The
light-matter interaction will be described in the velocity gauge. The
hexagonal lattice Fig. 1(a) is spanned by the basis vectors: $\mathbf{a}%
_{1}=\left( \sqrt{3}a/2,a/2\right) $ and $\mathbf{a}_{2}=\left( \sqrt{3}%
a/2,-a/2\right) $, with the lattice spacing $a$. \ In reciprocal space, one
can choose the hexagonal or rhombic Brillouin zone. For integration, it is
convenient to choose the rhombic Brillouin zone shown in Fig. 1(b) formed by
two vectors $\mathbf{b}_{1}$ and $\mathbf{b}_{2}$, with the reciprocal
lattice spacing - $k_{b}=4\pi /\sqrt{3}a$. 
\begin{figure}[tbp]
\includegraphics[width=.48\textwidth]{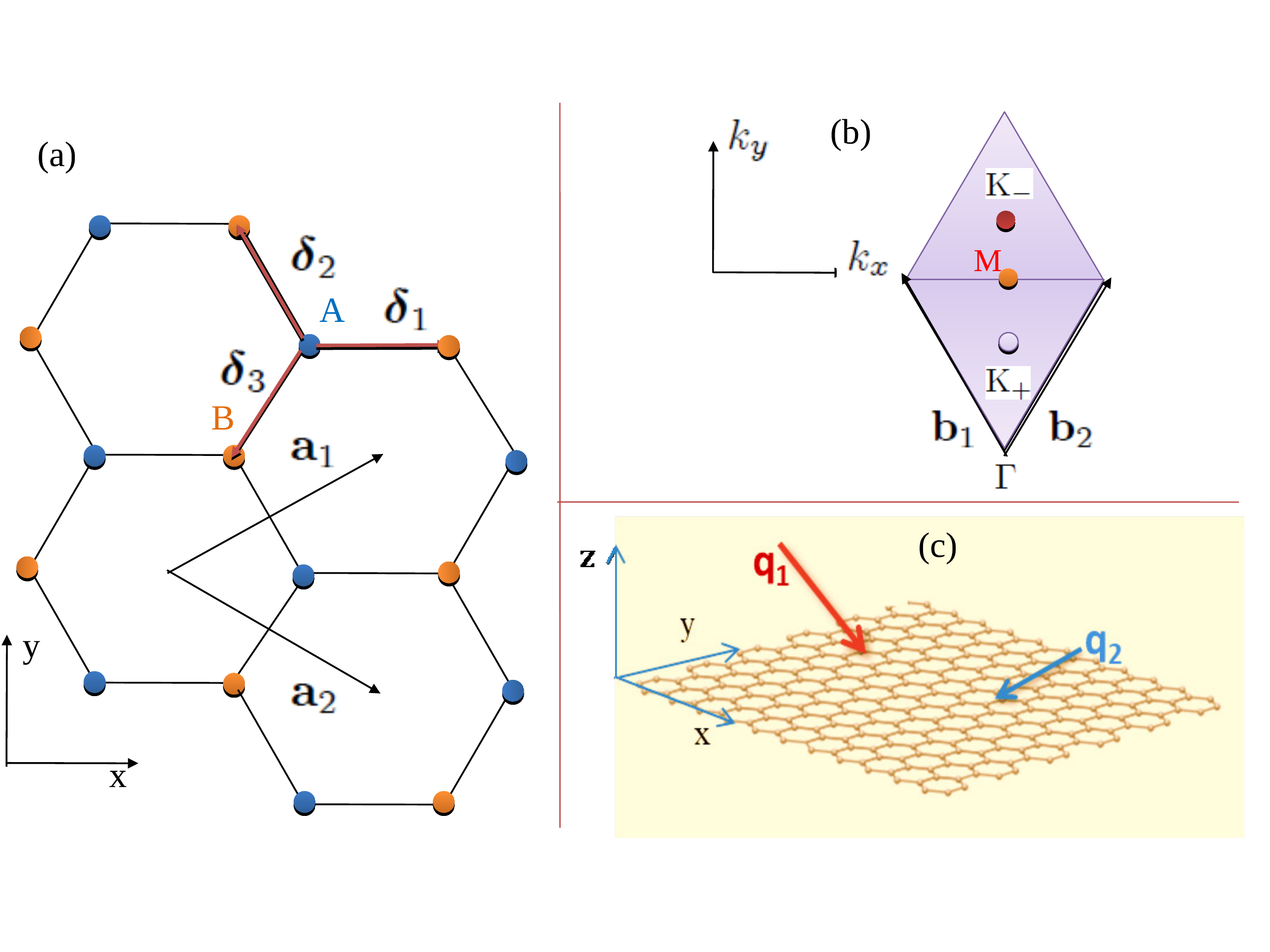}
\caption{(a) Hexagonal lattice in real space with two sublattices, A and B.
The vectors $\mbox{\boldmath{$\delta$}}_{1}$, $\mbox{\boldmath{$\delta$}}%
_{2} $, and $\mbox{\boldmath{$\delta$}}_{3}$ connect nearest neighbor atoms.
The vectors $\mathbf{a}_{1}=\mbox{\boldmath{$\delta$}}_{1}- 
\mbox{\boldmath{$\delta$}}_{3}$ and $\mathbf{a}_{2}=\mbox{\boldmath{$%
\delta$}}_{1}- \mbox{\boldmath{$\delta$}}_{2}$ are the basis vectors. (b)
The rhombical first Brillouin zone of reciprocal lattice with basis vectors $%
\mathbf{b}_{1}=\left( -2\protect\pi /\left( a\protect\sqrt{3}\right) ,2%
\protect\pi /a\right) $ and $\mathbf{b}_{2}=\left( 2\protect\pi /\left( a%
\protect\sqrt{3}\right) ,2\protect\pi /a\right) $. (c) A sketch of the
interaction geometry with obliquely incident waves.}
\end{figure}
The tight-binding Hamiltonian in the first nearest-neighbor approximation
can be written as 
\begin{equation}
\widehat{H}_{0}=-\gamma _{0}\sum_{\left\langle i,j\right\rangle
s_{z}}c_{is_{z}}^{\dagger }c_{js_{z}},  \label{Hm}
\end{equation}%
where $c_{is_{z}}^{\dagger }$ creates an electron with spin polarization $%
s_{z}$ at site $i$, and $\left\langle i,j\right\rangle $ runs over all the
first nearest-neighbor hopping sites with the transfer energy $\gamma _{0}$.
By performing Fourier transformations and choosing the basis $\left\{
|A\rangle ,|B\rangle \right\} $ $\otimes $ $\left\{ |\uparrow \rangle
,|\downarrow \rangle \right\} $, from Eq. (\ref{Hm}) one can obtain the
Hamiltonian%
\begin{equation}
\widehat{H}_{0}\left( \mathbf{k}\right) =\left[ 
\begin{array}{cc}
0 & -\gamma _{0}f\left( \mathbf{k}\right) \\ 
-\gamma _{0}f^{\ast }\left( \mathbf{k}\right) & 0%
\end{array}%
\right] ,  \label{H(k)}
\end{equation}%
where%
\begin{equation}
f\left( \mathbf{k}\right) =\sum_{i=1}^{3}\exp (i\mathbf{k}\cdot %
\mbox{\boldmath{$\delta$}}_{i})=e^{i\frac{ak_{x}}{\sqrt{3}}}+2e^{-i\frac{%
ak_{x}}{2\sqrt{3}}}\cos \left( \frac{ak_{y}}{2}\right) .  \label{f(k)}
\end{equation}%
Note that near the two Dirac points $\gamma _{0}f\left( \mathbf{k}\right)
=\hbar \mathrm{v}_{F}\left( ik_{x}\mp k_{y}\right) $, where $\mathrm{v}_{F}=%
\sqrt{3}a\gamma _{0}/2\hbar $ is the Fermi velocity. The spin $s_{z}=\pm 1$
is a good quantum number. For the issue considered, there are no spin-flip
transitions, and the spin index $s_{z}$ can be considered as a parameter.

In the presence of the radiation field with the vector potential $\mathbf{A}$
the Hamiltonian is obtained by Peierls substitution, i.\thinspace
e.\thinspace , $\mathbf{k}\rightarrow \mathbf{k}+e\mathbf{A/(\hbar }c\mathbf{%
)}$, where $\hbar $ is the Planck's constant, $e$ is the elementary charge, $%
c$ is the light speed in a vacuum. \ In general, one should keep all orders
of $\mathbf{A}$ in $\widehat{H}_{0}\left( \mathbf{k}+e\mathbf{A/(\hbar }c%
\mathbf{)}\right) $\textbf{. }For the second-order processes expanding the
Hamiltonian to the second order in the vector potential, one has 
\begin{equation}
\widehat{H}_{\mathrm{int}}=\frac{e}{\,c}A_{\alpha }\widehat{\mathrm{v}}%
_{\alpha }+\frac{e^{2}}{c^{2}}A_{\alpha }A_{\beta }\widehat{\tau }_{\alpha
\beta },  \label{Hint}
\end{equation}%
where 
\begin{equation}
\widehat{\mathrm{v}}_{\alpha }=\frac{1}{\hbar \,}\frac{\partial \widehat{H}%
_{0}\left( \mathbf{k}\right) }{\partial k_{\alpha }}  \label{va}
\end{equation}%
is the velocity operator and 
\begin{equation}
\widehat{\tau }_{\alpha \beta }=\frac{1}{\hbar \,^{2}}\frac{\partial ^{2}%
\widehat{H}_{0}\left( \mathbf{k}\right) }{\partial k_{\alpha }\partial
k_{\beta }}  \label{tab}
\end{equation}%
is the stress tensor operator. Hereafter summation over the repeated greek
indices is implied. The second term in Eq. (\ref{Hint}) is the diamagnetic
term\cite{Gusynin} that is absent in the DCA: $\widehat{\tau }_{\alpha \beta
}=0$. Note that within this tight-binding model there is no contribution
arising from the $z$ component of the vector potential. The latter describes
the in-plane magnetic field or out of the plane electric field component.%
\textrm{\ }Conduction and valence bands of hexagonal two-dimensional
nanostructures are formed from $\pi $ orbitals of material atoms. As a
result, the transitions in $z$ polarization are possible only with allowance
for other, $\sigma $ orbitals. These orbitals are separated from $\pi $
orbitals by a large energy gap and these transitions can be neglected.

Taking into account Eqs. (\ref{H(k)}) and (\ref{f(k)}) the velocity operator
(\ref{va}) can be represented as%
\begin{equation}
\widehat{\mathbf{v}}\left( \mathbf{k}\right) =\mathrm{v}_{F}\left[ 
\begin{array}{cc}
0 & \mathbf{\Lambda }\left( \mathbf{k}\right) \\ 
\mathbf{\Lambda }^{\ast }\left( \mathbf{k}\right) & 0%
\end{array}%
\right] ,  \label{velop}
\end{equation}%
where 
\begin{equation}
\Lambda _{x}=-\frac{2}{3}i\left( e^{i\frac{ak_{x}}{\sqrt{3}}}-e^{-i\frac{%
ak_{x}}{2\sqrt{3}}}\cos \left( \frac{ak_{y}}{2}\right) \right) ,  \label{vx}
\end{equation}%
\begin{equation}
\Lambda _{y}=\frac{2}{\sqrt{3}}e^{-i\frac{ak_{x}}{2\sqrt{3}}}\sin \left( 
\frac{ak_{y}}{2}\right) .  \label{vy}
\end{equation}%
Similarly the stress tensor operator (\ref{tab}) will be 
\begin{equation}
\widehat{\tau }_{\alpha \beta }=\frac{\mathrm{v}_{F}^{2}}{\gamma _{0}}\left[ 
\begin{array}{cc}
0 & \Upsilon _{\alpha \beta } \\ 
\Upsilon _{\alpha \beta }^{\ast } & 0%
\end{array}%
\right] ,  \label{tauab}
\end{equation}%
where 
\begin{equation*}
\Upsilon _{xx}=\frac{4}{9}\left[ e^{i\frac{ak_{x}}{\sqrt{3}}}+\frac{1}{2}%
e^{-i\frac{ak_{x}}{2\sqrt{3}}}\cos \left( \frac{ak_{y}}{2}\right) \right] ,
\end{equation*}%
\begin{equation*}
\Upsilon _{yy}=\frac{2}{3}e^{-i\frac{ak_{x}}{2\sqrt{3}}}\cos \left( \frac{%
ak_{y}}{2}\right) ,
\end{equation*}%
\begin{equation*}
\Upsilon _{xy}=\Upsilon _{yx}=-\frac{2i}{3\sqrt{3}}e^{-i\frac{ak_{x}}{2\sqrt{%
3}}}\sin \left( \frac{ak_{y}}{2}\right) .
\end{equation*}%
The vector potential is assumed to be%
\begin{equation}
\mathbf{A}\left( \mathbf{r},t\right) \mathbf{=}\sum_{\delta ,s=\pm }\mathbf{A%
}(s\omega _{\delta })e^{is\left( \mathbf{q}_{\delta }\cdot \mathbf{r}-\omega
_{\delta }t\right) };\ \mathbf{A}(-\omega _{\delta })=\mathbf{A}^{\ast
}(\omega _{\delta }),  \label{VP}
\end{equation}%
where summation is over involved frequencies. The interaction Hamiltonian
can be written as 
\begin{equation}
\widehat{H}_{\mathrm{int}}=\widehat{H}_{\mathrm{int}}^{(1)}+\widehat{H}_{%
\mathrm{int}}^{(2)},  \label{intH}
\end{equation}%
where 
\begin{equation}
\widehat{H}_{\mathrm{int}}^{(1)}=\frac{e}{c}\sum\limits_{s,\delta }\widehat{%
\mathrm{v}}_{\eta }A_{\eta }(s\omega _{\delta })\exp \left[ is\left( \mathbf{%
q}_{\delta }\cdot \mathbf{r}-\omega _{\delta }t\right) \right] ,
\label{int1}
\end{equation}%
and%
\begin{equation*}
\widehat{H}_{\mathrm{int}}^{(2)}=\frac{e^{2}}{c^{2}}\sum\limits_{s,\delta
}\sum\limits_{s_{1},\delta _{1}}A_{\eta }(s\omega _{\delta })A_{\beta
}(s_{1}\omega _{\delta _{1}})\widehat{\tau }_{\eta \beta }
\end{equation*}%
\begin{equation}
\times \exp \left[ i\left( s_{1}\mathbf{q}_{\delta _{1}}+s\mathbf{q}_{\delta
}\right) \mathbf{r}-i\left( s_{1}\omega _{\delta _{1}}+s\omega _{\delta
}\right) t\right] .  \label{int2}
\end{equation}%
Here $\mathbf{q}_{\delta }$ is the in plane wave vector\textbf{.}

The eigenstates of the Hamiltonian (\ref{H(k)}) with the combined quantum
number $m=\{s_{m},\mathbf{k}_{m}\}$ are:%
\begin{equation}
|m\rangle =|s_{m},\mathbf{k}_{m}\rangle e^{i\mathbf{k}_{m}\mathbf{r}},
\label{free}
\end{equation}%
where%
\begin{equation}
|s_{m},\mathbf{k}_{m}\rangle =\frac{1}{\sqrt{2}}\left[ 
\begin{array}{c}
e^{i\Theta \left( \mathbf{k}_{m}\right) } \\ 
s_{m}%
\end{array}%
\right]  \label{bf}
\end{equation}%
are spinors corresponding to energies%
\begin{equation}
\mathcal{E}\left( m\right) =s_{m}\gamma _{0}\left\vert f\left( \mathbf{k}%
_{m}\right) \right\vert .  \label{energy}
\end{equation}%
The band index $s_{m}=\pm 1$: for conduction ($s_{m}=1$) and valence ($%
s_{m}=-1$) bands, and $\Theta \left( \mathbf{k}_{m}\right) =\arg \left(
-\gamma _{0}f\left( \mathbf{k}_{m}\right) \right) $.

In order to develop a microscopic theory of the nonlinear interaction of a
nanostructure with a multicolor radiation field we need to solve the master
equation for the density matrix $\rho _{mn}$:%
\begin{equation*}
i\hbar \frac{\partial \rho _{mn}}{\partial t}=(\mathcal{E}\left( m\right) -%
\mathcal{E}\left( n\right) )\rho _{mn}
\end{equation*}%
\begin{equation}
+\sum_{l}\left[ \langle m|\widehat{H}_{\mathrm{int}}|l\rangle \rho
_{ln}-\rho _{ml}\langle l|\widehat{H}_{\mathrm{int}}|n\rangle \right]
-i\hbar \gamma \left( \rho _{mn}-\rho _{mn}^{(0)}\right) ,  \label{main}
\end{equation}%
where $\rho _{mn}^{(0)}$ is the equilibrium density matrix to which the
system relaxes at a rate $\gamma $. We construct $\rho _{mn}^{(0)}$ from the
filling of electron states according to the Fermi--Dirac-distribution:%
\begin{equation*}
\rho _{mn}^{(0)}=n_{F}\left( m\right) \delta _{mn},
\end{equation*}%
where 
\begin{equation}
n_{F}\left( m\right) \equiv n_{F}\left( s_{m},\mathbf{k}_{m}\right) =\frac{1%
}{1+\exp \left( \frac{s_{m}\gamma _{0}\left\vert f\left( \mathbf{k}%
_{m}\right) \right\vert -\varepsilon _{F}}{k_{B}T}\right) }.  \label{fd}
\end{equation}%
Here $\varepsilon _{F}$ is the Fermi energy, $k_{B}$ is Boltzmann's
constant, and $T$ is the absolute temperature. Note that this relaxation
approximation provides an accurate description for optical field components
oscillating at frequencies $\omega >>\gamma $.

We will solve Eq. (\ref{main}) in the scope of perturbation theory:%
\begin{equation}
\rho _{mn}\left( t\right) =\rho _{mn}^{(0)}+\rho _{mn}^{(1)}\left( t\right)
+\rho _{mn}^{(2)}\left( t\right) +^{_{...}}.  \label{pert}
\end{equation}%
From Eq. (\ref{main}) we have the following equations for $\rho
_{mn}^{(1)}\left( t\right) \sim A$, and $\rho _{mn}^{(2)}\left( t\right)
\sim A^{2}$:%
\begin{equation*}
i\hbar \frac{\partial \rho _{mn}^{(1)}\left( t\right) }{\partial t}=\left( 
\mathcal{E}\left( m\right) -\mathcal{E}\left( n\right) -i\hbar \gamma
\right) \rho _{mn}^{(1)}\left( t\right)
\end{equation*}%
\begin{equation}
+\sum_{l}\left[ \langle m|\widehat{H}_{\mathrm{int}}^{(1)}|l\rangle \rho
_{ln}^{(0)}-\rho _{ml}^{(0)}\langle l|\widehat{H}_{\mathrm{int}%
}^{(1)}|n\rangle \right] ,  \label{p2}
\end{equation}%
\begin{equation*}
i\hbar \frac{\partial \rho _{mn}^{(2)}\left( t\right) }{\partial t}=\left( 
\mathcal{E}\left( m\right) -\mathcal{E}\left( n\right) -i\hbar \gamma
\right) \rho _{mn}^{(2)}\left( t\right)
\end{equation*}%
\begin{equation*}
+\sum_{l}\left[ \langle m|\widehat{H}_{\mathrm{int}}^{(1)}|l\rangle \rho
_{ln}^{(1)}\left( t\right) -\rho _{ml}^{(1)}\left( t\right) \langle l|%
\widehat{H}_{\mathrm{int}}^{(1)}|n\rangle \right]
\end{equation*}%
\begin{equation}
+\sum_{l}\left[ \langle m|\widehat{H}_{\mathrm{int}}^{(2)}|l\rangle \rho
_{ln}^{(0)}-\rho _{ml}^{(0)}\langle l|\widehat{H}_{\mathrm{int}%
}^{(2)}|n\rangle \right] .  \label{p3}
\end{equation}%
The solutions to Eqs. (\ref{p2}) and (\ref{p3}) are%
\begin{widetext}
\begin{equation}
\rho _{mn}^{(1)}=\frac{e}{c}\sum\limits_{s,\delta }\frac{A_{\eta }(s\omega
_{\delta })\exp \left( -is\omega _{\delta }t\right) \langle m|\widehat{%
\mathrm{v}}_{\eta }e^{is\mathbf{q}_{\delta }\mathbf{r}}|n\rangle }{\mathcal{E%
}\left( m\right) -\mathcal{E}\left( n\right) -s\hbar \omega _{\delta
}-i\hbar \gamma }\left( n_{F}\left( m\right) -n_{F}\left( n\right) \right) ,
\label{sp2}
\end{equation}%
\begin{equation*}
\rho _{mn}^{(2)}=\frac{e^{2}}{c^{2}}\sum_{l}\sum\limits_{s,\delta
}\sum\limits_{s_{1},\delta _{1}}\frac{A_{\beta }(s_{1}\omega _{\delta
_{1}})A_{\eta }(s\omega _{\delta })\exp \left( -i\left( s_{1}\omega _{\delta
_{1}}+s\omega _{\delta }\right) t\right) }{\mathcal{E}\left( m\right) -%
\mathcal{E}\left( n\right) -\hbar \left( s_{1}\omega _{\delta _{1}}+s\omega
_{\delta }\right) -i\hbar \gamma }
\end{equation*}%
\begin{equation*}
\times \left[ \frac{\langle m|\widehat{\mathrm{v}}_{\beta }e^{is_{1}\mathbf{q%
}_{\delta _{1}}\mathbf{r}}|l\rangle \langle l|\widehat{\mathrm{v}}_{\eta
}e^{is\mathbf{q}_{\delta }\mathbf{r}}|n\rangle }{\mathcal{E}\left( m\right) -%
\mathcal{E}\left( l\right) -s_{1}\hbar \omega _{\delta _{1}}-i\hbar \gamma }%
\left( n_{F}\left( m\right) -n_{F}\left( l\right) \right) -\frac{\langle m|%
\widehat{\mathrm{v}}_{\eta }e^{is\mathbf{q}_{\delta }\mathbf{r}}|l\rangle
\langle l|\widehat{\mathrm{v}}_{\beta }e^{is_{1}\mathbf{q}_{\delta _{1}}%
\mathbf{r}}|n\rangle }{\mathcal{E}\left( l\right) -\mathcal{E}\left(
n\right) -s_{1}\hbar \omega _{\delta _{1}}-i\hbar \gamma }\left( n_{F}\left(
l\right) -n_{F}\left( n\right) \right) \right]
\end{equation*}%
\begin{equation}
+\frac{e^{2}}{c^{2}}\sum\limits_{s,\delta }\sum\limits_{s_{1},\delta _{1}}%
\frac{A_{\eta }(s\omega _{\delta })A_{\beta }(s_{1}\omega _{\delta
_{1}})\exp \left( -i\left( s_{1}\omega _{\delta _{1}}+s\omega _{\delta
}\right) t\right) \langle m|\widehat{\tau }_{\eta \beta }e^{is_{1}\mathbf{q}%
_{\delta _{1}}\mathbf{r}}e^{is\mathbf{q}_{\delta }\mathbf{r}}|n\rangle }{%
\mathcal{E}\left( m\right) -\mathcal{E}\left( n\right) -\hbar \left(
s_{1}\omega _{\delta _{1}}+s\omega _{\delta }\right) -i\hbar \gamma }\left(
n_{F}\left( m\right) -n_{F}\left( n\right) \right) .  \label{sp3}
\end{equation}%
With the help of solution (\ref{sp3}) one can calculate physical observables
to investigate second order nonlinear response of 2D nanostructures. The
transition matrix elements for velocity (\ref{velop}) and stress tensor (\ref%
{tauab}) operators can be calculated with the help of Eqs. (\ref{velop}), (%
\ref{tauab}), (\ref{free}), and (\ref{bf}). As a result we obtain%
\begin{equation}
\langle n|\hat{\mathrm{v}}_{\alpha }e^{i\mathbf{q}\cdot \mathbf{r}}|m\rangle
=\langle s_{n},\mathbf{k}_{n}|\hat{\mathrm{v}}_{\alpha }|s_{m},\mathbf{k}%
_{m}\rangle \left( 2\pi \right) ^{2}\delta \left( \mathbf{k}_{m}+\mathbf{q-k}%
_{n}\right) ,  \label{vnmq}
\end{equation}%
\begin{equation}
\langle n|\widehat{\tau }_{\alpha \beta }e^{i\mathbf{q}\cdot \mathbf{r}%
}|m\rangle =\langle s_{n},\mathbf{k}_{n}|\widehat{\tau }_{\alpha \beta
}|s_{m},\mathbf{k}_{m}\rangle \left( 2\pi \right) ^{2}\delta \left( \mathbf{k%
}_{m}+\mathbf{q-k}_{n}\right) ,  \label{tnmq}
\end{equation}%
where 
\begin{equation}
\langle s_{n},\mathbf{k}_{n}|\hat{\mathrm{v}}_{\alpha }|s_{m},\mathbf{k}%
_{m}\rangle =\frac{\mathrm{v}_{F}}{2}\left[ s_{m}\mathbf{\Lambda }\left( 
\mathbf{k}_{m}\right) e^{-i\Theta \left( \mathbf{k}_{n}\right) }+s_{n}%
\mathbf{\Lambda }^{\ast }\left( \mathbf{k}_{m}\right) e^{i\Theta \left( 
\mathbf{k}_{m}\right) }\right] .  \label{vnm}
\end{equation}%
and

\begin{equation}
\langle s_{n},\mathbf{k}_{n}|\widehat{\tau }_{\alpha \beta }|s_{m},\mathbf{k}%
_{m}\rangle =\frac{\mathrm{v}_{F}^{2}}{2\gamma _{0}}\left[ s_{m}\Upsilon
_{\alpha \beta }\left( \mathbf{k}_{m}\right) e^{-i\Theta \left( \mathbf{k}%
_{n}\right) }+s_{n}\Upsilon _{\alpha \beta }^{\ast }\left( \mathbf{k}%
_{m}\right) e^{i\Theta \left( \mathbf{k}_{m}\right) }\right] .  \label{tnm}
\end{equation}%
The Dirac delta function in Eqs. (\ref{vnmq}) and (\ref{tnmq}) expresses
conservation law for momentum.

\section{Second order nonlinear response of 2D hexagonal nanostructure}

With the help of solutions (\ref{sp2}) and (\ref{sp3}) of the quantum master
equation (\ref{main}), obtained in the previous section, one can investigate
the linear and second-order nonlinear electromagnetic response of hexagonal
nanostructure. The linear response beyond the Dirac cone approximation is
well investigated\cite{Nov2} and we will concentrate on the second-order
nonlinear electromagnetic response. Along with the graphene we will present
the results for silicene. Germanene and stanene have parameters close to
silicene and the results for these materials will be almost identical. We
will consider the spectral range when the Brillouin zone of a hexagonal
tight-binding nanostructure is excited out of Dirac two cones. First, we
calculate the second-order conductivity tensor. The total current density
operator is obtained by differentiating $\widehat{H}_{\mathrm{int}}$ with
respect to $A_{\alpha }$, 
\begin{equation}
\widehat{j}_{\alpha }=-\hbar \frac{\partial \widehat{H}_{\mathrm{int}}}{%
\partial \left( A_{\alpha }/c\right) }=\widehat{j}_{\alpha }^{p}+\widehat{j}%
_{\alpha }^{d},  \label{cr1}
\end{equation}%
and consists of the usual paramagnetic part 
\begin{equation}
\widehat{j}_{\alpha }^{p}=-e\widehat{\mathrm{v}}_{\alpha }  \label{cr2}
\end{equation}%
and diamagnetic part 
\begin{equation}
\widehat{j}_{\alpha }^{d}=-\frac{e^{2}}{c}A_{\beta }\widehat{\tau }_{\alpha
\beta }.  \label{cr3}
\end{equation}%
The total current can be written as 
\begin{equation*}
j_{\alpha }(t,\mathbf{r})=g_{s}\sum_{mn}\langle s_{n},\mathbf{k}_{n}|%
\widehat{j}_{\alpha }|s_{m},\mathbf{k}_{m}\rangle e^{i\left( \mathbf{k}_{m}-%
\mathbf{k}_{n}\right) \mathbf{r}}\rho _{mn}(t),
\end{equation*}%
where $g_{s}=2$\ is the spin degeneracy factor. For the second-order
nonlinear current we will have 
\begin{equation*}
j_{\alpha }^{(2)}(t,\mathbf{r})=-g_{s}e\sum_{mn}\langle s_{n},\mathbf{k}_{n}|%
\widehat{\mathrm{v}}_{\alpha }|s_{m},\mathbf{k}_{m}\rangle e^{i\left( 
\mathbf{k}_{m}-\mathbf{k}_{n}\right) \mathbf{r}}\rho _{mn}^{(2)}(t)
\end{equation*}%
\begin{equation}
-g_{s}\frac{e^{2}}{c}\sum_{mn}\langle s_{n},\mathbf{k}_{n}|\widehat{\tau }%
_{\alpha \beta }|s_{m},\mathbf{k}_{m}\rangle A_{\beta }e^{i\left( \mathbf{k}%
_{m}-\mathbf{k}_{n}\right) \mathbf{r}}\rho _{mn}^{(1)}(t)  \label{tc2}
\end{equation}%
Taking into account the relation 
\begin{equation}
j_{\alpha }^{(2)}(t,\mathbf{r})=j_{\alpha }^{(2)}(\omega ,\mathbf{q}%
)e^{i\left( \mathbf{q}\cdot \mathbf{r}-\omega t\right) }+c.c.,  \label{cur}
\end{equation}%
we introduce conductivity tensor $\sigma _{\alpha \beta \eta }$ via the
electrical field strength Fourier amplitudes:

\begin{equation*}
j_{\alpha }^{(2)}\left( \omega _{3},\mathbf{q}_{3}\right) =g_{\omega }\sigma
_{\alpha \beta \eta }\left( \omega _{3},\mathbf{q}_{3};\omega _{1},\mathbf{q}%
_{1},\omega _{2},\mathbf{q}_{2}\right) E_{\beta }\left( \omega _{1}\right)
E_{\eta }\left( \omega _{2}\right) 
\end{equation*}%
\begin{equation}
=g_{\omega }\sigma _{\alpha \beta \eta }\left( \omega _{3},\mathbf{q}%
_{3};\omega _{1},\mathbf{q}_{1},\omega _{2},\mathbf{q}_{2}\right) \frac{%
\omega _{1}\omega _{2}}{c^{2}}A_{\beta }\left( \omega _{1}\right) A_{\eta
}\left( \omega _{2}\right) ,  \label{cur2}
\end{equation}%
where $g_{\omega }$ is the degeneracy factor. For single pump wave, we have $%
g_{\omega }=1/2$, otherwise $g_{\omega }=1$. For the physical reasons, we
separate paramagnetic and diamagnetic parts of the conductivity tensor: 
\begin{equation*}
\sigma _{\alpha \beta \eta }\left( \omega _{3},\mathbf{q}_{3};\omega _{1},%
\mathbf{q}_{1},\omega _{2},\mathbf{q}_{2}\right) =\sigma _{\alpha \beta \eta
}^{(p)}\left( \omega _{3},\mathbf{q}_{3};\omega _{1},\mathbf{q}_{1},\omega
_{2},\mathbf{q}_{2}\right) +\sigma _{\alpha \beta \eta }^{(d)}\left( \omega
_{3},\mathbf{q}_{3};\omega _{1},\mathbf{q}_{1},\omega _{2},\mathbf{q}%
_{2}\right) .
\end{equation*}%
From Eqs. (\ref{sp3}), (\ref{tc2}), (\ref{cur}) and (\ref{cur2}) for the
both paramagnetic and diamagnetic parts of the nonlinear conductivity tensor
we will have 
\begin{equation}
\sigma _{\alpha \beta \eta }^{(p,d)}\left( \omega _{3},\mathbf{q}_{3};\omega
_{1},\mathbf{q}_{1},\omega _{2},\mathbf{q}_{2}\right) =F_{\alpha \beta \eta
}^{(p,d)}\left( \omega _{1},\mathbf{q}_{1},\omega _{2},\mathbf{q}_{2}\right)
+F_{\alpha \eta \beta }^{(p,d)}\left( \omega _{2},\mathbf{q}_{2},\omega _{1},%
\mathbf{q}_{1}\right) ,  \label{second}
\end{equation}%
where%
\begin{equation*}
F_{\alpha \beta \eta }^{(p)}\left( \omega _{1},\mathbf{q}_{1},\omega _{2},%
\mathbf{q}_{2}\right) =-\frac{2e^{3}}{\omega _{1}\omega _{2}}\frac{1}{\left(
2\pi \right) ^{2}}\sum_{s_{m},s_{n,}s_{l}}\int_{BZ}d\mathbf{k}\frac{\langle
s_{n},\mathbf{k}-\mathbf{q}_{2}|\hat{\mathrm{v}}_{\alpha }|s_{m},\mathbf{k}+%
\mathbf{q}_{1}\rangle \langle s_{m},\mathbf{k}+\mathbf{q}_{1}|\hat{\mathrm{v}%
}_{\beta }|s_{l},\mathbf{k}\rangle \langle s_{l},\mathbf{k}|\hat{\mathrm{v}}%
_{\eta }|s_{n},\mathbf{k}-\mathbf{q}_{2}\rangle }{s_{m}\gamma _{0}\left\vert
f\left( \mathbf{k}+\mathbf{q}_{1}\right) \right\vert -s_{n}\gamma
_{0}\left\vert f\left( \mathbf{k}-\mathbf{q}_{2}\right) \right\vert -\hbar
\left( \omega _{1}+\omega _{2}\right) -i\hbar \gamma }
\end{equation*}%
\begin{equation}
\times \left[ \frac{n_{F}\left( s_{m},\mathbf{k}+\mathbf{q}_{1}\right)
-n_{F}\left( s_{l},\mathbf{k}\right) }{s_{m}\gamma _{0}\left\vert f\left( 
\mathbf{k}+\mathbf{q}_{1}\right) \right\vert -s_{l}\gamma _{0}\left\vert
f\left( \mathbf{k}\right) \right\vert -\hbar \omega _{1}-i\hbar \gamma }-%
\frac{n_{F}\left( s_{l},\mathbf{k}\right) -n_{F}\left( s_{n},\mathbf{k}-%
\mathbf{q}_{2}\right) }{s_{l}\gamma _{0}\left\vert f\left( \mathbf{k}\right)
\right\vert -s_{n}\gamma _{0}\left\vert f\left( \mathbf{k}-\mathbf{q}%
_{2}\right) \right\vert -\hbar \omega _{2}-i\hbar \gamma )}\right] .
\label{Fp}
\end{equation}%
and%
\begin{equation*}
F_{\alpha \beta \eta }^{(d)}\left( \omega _{1},\mathbf{q}_{1},\omega _{2},%
\mathbf{q}_{2}\right) =-\frac{2e^{3}}{\omega _{1}\omega _{2}}\frac{1}{\left(
2\pi \right) ^{2}}\sum_{s_{m},s_{n}}\int_{BZ}d\mathbf{k}\left[ \frac{\langle
s_{n},\mathbf{k}|\widehat{\mathrm{v}}_{\alpha }|s_{m},\mathbf{k}+\mathbf{q}%
_{3}\rangle \langle s_{m},\mathbf{k}+\mathbf{q}_{3}|\widehat{\tau }_{\beta
\eta }|s_{n},\mathbf{k}\rangle }{\mathcal{E}\left( s_{m},\mathbf{k}+\mathbf{q%
}_{3}\right) -\mathcal{E}\left( s_{n},\mathbf{k}\right) -\hbar \omega
_{3}-i\hbar \gamma }\right. 
\end{equation*}%
\begin{equation}
\times \left( n_{F}\left( s_{m},\mathbf{k}+\mathbf{q}_{3}\right)
-n_{F}\left( s_{n},\mathbf{k}\right) \right) +\left. \frac{\langle s_{m},%
\mathbf{k}+\mathbf{q}_{1}|\widehat{\mathrm{v}}_{\eta }|s_{n},\mathbf{k}%
\rangle \langle s_{n},\mathbf{k}|\widehat{\tau }_{\alpha \beta }|s_{m},%
\mathbf{k}+\mathbf{q}_{1}\rangle }{\mathcal{E}\left( s_{m},\mathbf{k}+%
\mathbf{q}_{1}\right) -\mathcal{E}\left( s_{n},\mathbf{k}\right) -\hbar
\omega _{1}-i\hbar \gamma }\left( n_{F}\left( s_{m},\mathbf{k}+\mathbf{q}%
_{1}\right) -n_{F}\left( s_{n},\mathbf{k}\right) \right) \right] .
\label{Fd}
\end{equation}%
\end{widetext}In the language of Feynman diagrams paramagnetic and
diamagnetic parts correspond to triangular and nonlinear bubble diagrams,%
\cite{diagr} respectively. As is seen from Eqs. (\ref{second}) the
conductivity tensor is symmetric in its components and arguments: 
\begin{equation}
\sigma _{\alpha \eta \beta }\left( \omega _{3},\mathbf{q}_{3};\omega _{2},%
\mathbf{q}_{2},\omega _{1},\mathbf{q}_{1}\right) =\sigma _{\alpha \beta \eta
}\left( \omega _{3},\mathbf{q}_{3};\omega _{1},\mathbf{q}_{1},\omega _{2},%
\mathbf{q}_{2}\right) .  \label{symetri}
\end{equation}%
Let us consider the second-order conductivity tensor given by Eq. (\ref%
{second}). Following convention,\cite{Boyd2003} we have written $\sigma
_{\alpha \beta \eta }\left( \omega _{3},\mathbf{q}_{3};\omega _{1},\mathbf{q}%
_{1},\omega _{2},\mathbf{q}_{2}\right) $ as a function of three frequencies
and wave vectors. The first two arguments are associated with the time-space
dependence of the resulting field $\exp (i\mathbf{q}_{3}\mathbf{r}-i\omega
_{3}t)$ and we have energy and momentum conservation: $\omega _{3}=\omega
_{1}+\omega _{2}$ and $\mathbf{q}_{3}=\mathbf{q}_{1}+\mathbf{q}_{2}$ at the
three wave mixing. Thus, we have mutual interaction of three waves and for a
complete description of the interaction of these waves we need to determine
the tensors $\sigma _{\alpha \beta \eta }\left( \omega _{3},\mathbf{q}%
_{3};\omega _{1},\mathbf{q}_{1},\omega _{2},\mathbf{q}_{2}\right) $, $\sigma
_{\alpha \beta \eta }\left( \omega _{1},\mathbf{q}_{1};\omega _{3},\mathbf{q}%
_{3},-\omega _{2},-\mathbf{q}_{2}\right) $, and $\sigma _{\alpha \beta \eta
}\left( \omega _{2},\mathbf{q}_{2};\omega _{3},\mathbf{q}_{3}-\omega _{1},-%
\mathbf{q}_{1}\right) $, wherein we have two independent frequencies and
wavevectors. In particular $\sigma _{\alpha \beta \eta }\left( \omega _{3},%
\mathbf{q}_{3};\omega _{1},\mathbf{q}_{1},\omega _{2},\mathbf{q}_{2}\right) $
is responsible for the sum-frequency generation. At $\omega _{1}=\omega
_{2}\equiv \omega $ we have the second harmonic generation process. The
tensor $\sigma _{\alpha \beta \eta }\left( \omega _{1},\mathbf{q}_{1};\omega
_{3},\mathbf{q}_{3},-\omega _{2},-\mathbf{q}_{2}\right) $ is responsible for
the difference-frequency generation. In this case $\omega _{3}$ is known as
the pump frequency, $\omega _{2}$ the signal frequency, and $\omega _{1}$
the idler frequency. In the next, we will consider these processes
separately.

With the help of conductivity tensor in CGS units, one can calculate also
susceptibility tensor in SI units by the formula%
\begin{equation*}
\chi _{\alpha \beta \eta }\left( \omega _{3},\mathbf{q}_{3};\omega _{1},%
\mathbf{q}_{1},\omega _{2},\mathbf{q}_{2}\right)
\end{equation*}%
\begin{equation}
=\frac{4\pi i}{\omega _{3}}\sigma _{\alpha \beta \eta }\left( \omega _{3},%
\mathbf{q}_{3};\omega _{1},\mathbf{q}_{1},\omega _{2},\mathbf{q}_{2}\right) .
\label{CI2}
\end{equation}%
For easier comparison of the nonlinear response of the considered
nanostructure with known materials hereafter we will calculate
susceptibility tensor in SI units. For the final result Eqs. (\ref{Fp}) and (%
\ref{Fd}) should be integrated over the FBZ for a given geometry of the
incident fields, Fermi energy, and temperature. The paramagnetic part (\ref%
{Fp}) contains intraband contributions (two terms), as well as all types of
mixed interband and intraband contributions (six terms). The diamagnetic
part (\ref{Fd}) contains pure intraband and interband contributions. For
normal incidence ($\mathbf{q}_{1}=\mathbf{q}_{2}=0$) $\chi _{\alpha \beta
\eta }=0$ as expected from inversion symmetry of considered nanostructure.
For hexagonal nanostructure (Fig. 1(c)), we have two directions of interest
for in-plane wave vectors: along the zigzag direction or armchair one. As
expected from the symmetry, the results are identical. For concreteness, we
will direct all in-plane photon wave vectors along the $x$-axis (3D wave
vectors in the ZX plane). In this case nonzero components are $\chi _{xxx}$, 
$\chi _{xyy}$, $\chi _{yxy},\chi _{yyx}$. Here $\chi _{xxx}$, $\chi _{xyy}$
describe the generation of the p-polarized wave with p-polarized and
s-polarized waves, correspondingly. Then $\chi _{yxy}\ $and $\chi _{yyx}$
describe the generation of the s-polarized waves with mixed waves. Note that 
$\chi _{yxx}=0$, since p-polarized input waves can not generate s-polarized
output wave. Integration has been made over the rhombic Brillouin zone shown
in Fig. 1(b). In calculating the nonlinear susceptibility tensor, very dense 
$\mathbf{k}$\ mesh is needed. The convergence of $\mathbf{k}$\ mesh was
checked. The calculated nonlinear susceptibility tensor is converged with $%
2\times 10^{6}$\ final grid. All results have been calculated using uniform
mesh with $4\times 10^{6}$\ points. Also note that the obtained formulas are
not valid at the small frequencies $\omega \precsim \gamma $, and $\chi
_{\alpha \beta \eta }$ diverges at $\omega \rightarrow 0$. Besides, since we
adopted an independent quasiparticle picture, one should be also careful at
the applying obtained results to far off-resonant pump waves. At the
excitations of a nanostructure with the waves of photons energy $\hbar
\omega >>$ $\varepsilon _{F}$ one triggers photoexcitation cascade and, as a
result, the multiple hot carrier generation takes place in the nanostructure.%
\cite{hot} Meanwhile, near the Fermi level, these processes are suppressed
and we have the dominant contribution of pure optical transitions, thus Eq. (%
\ref{CI2}) can accurately describe second-order nonlinear optical response
for optical field components oscillating at the frequencies $\omega >\gamma $%
. 
\begin{figure}[tbp]
\includegraphics[width=.35\textwidth]{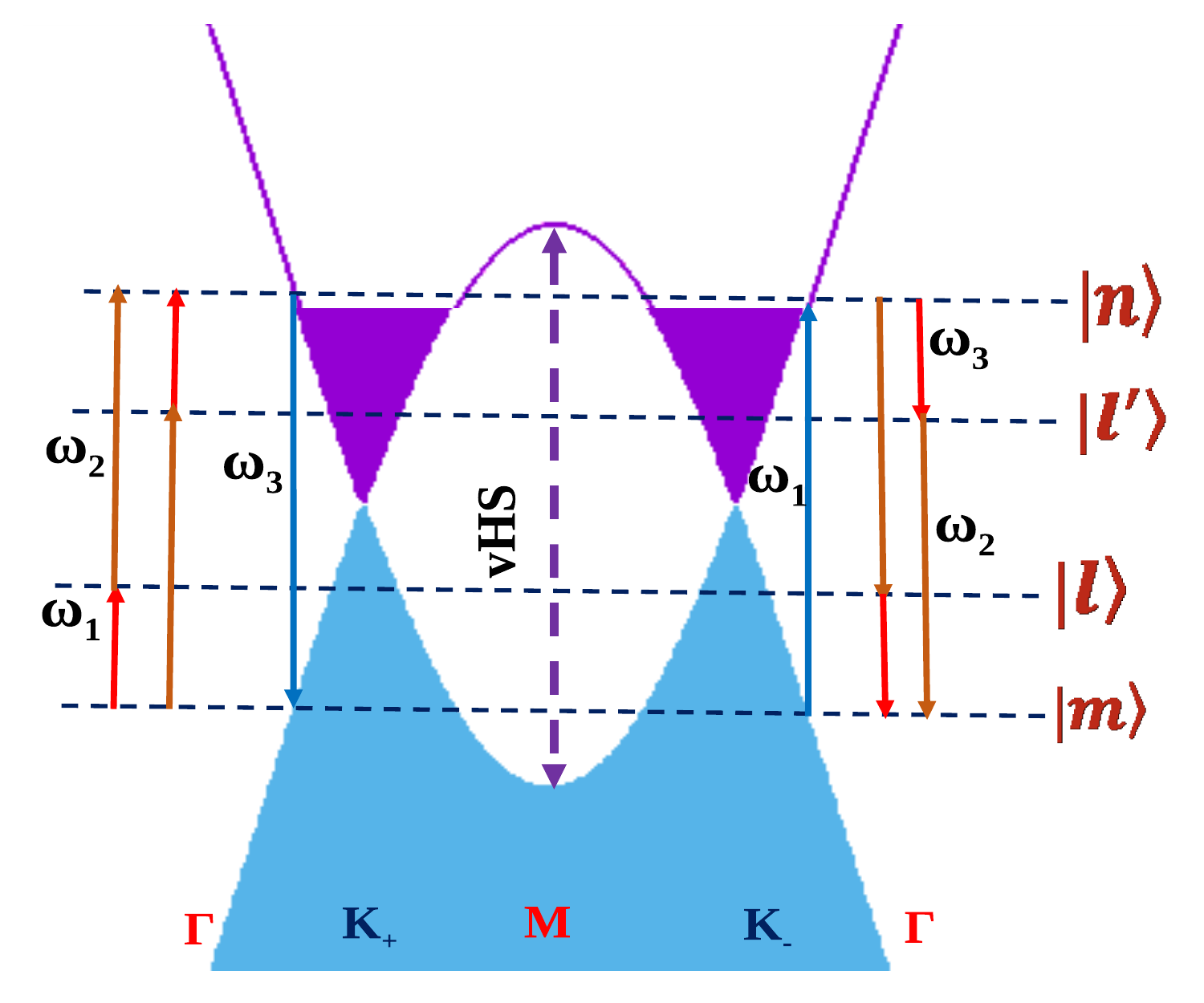}
\caption{Elementary three-wave mixing processes coupled to
interband/intraband transitions for sum (left arrows) and difference (right
arrows) frequency generation processes. It is shown dispersion relation
(cross-section $k_{x}=0$) where shading indicates filled electron states.
High-energy excitations are situated in the vicinity of the $\Gamma $\
point. Low-energy excitations are centered around the two points $K_{+}=k_{b}%
\widehat{\mathbf{y}}/\protect\sqrt{3}$\ and $K_{-}=2k_{b}\widehat{\mathbf{y}}%
/\protect\sqrt{3}$. Dashed arrow shows the optical resonance at the van Hove
singularity ($M=\protect\sqrt{3}k_{b}\widehat{\mathbf{y}}/2$). Diamagnetic
susceptibility (\protect\ref{Fd}) is a result of direct interband
transitions without intermediate states $|l\rangle $ and $|l^{\prime
}\rangle $.}
\end{figure}
It is clear that due to the electron-hole symmetry the absolute value of the
second-order susceptibility tensor is the same for $\pm \varepsilon _{F}$.
Thus, we will consider only the electron-doped system $\varepsilon _{F}>0$.
\ For high frequencies $\hbar \omega >$ $\varepsilon _{F}$, a purely
intraband contribution to the second-order susceptibility is very small, and
the three-wave mixing processes that give the main contribution to the
second-order susceptibility tensor are those in which the waves are coupled
to the interband transitions. The elementary three-wave-mixing resonant
processes coupled to the interband and intraband transitions are shown in
Fig. 2. Diamagnetic contribution to the susceptibility (\ref{Fd}) is
conditioned by the direct interband transitions without intermediate
intraband transitions. Meanwhile, paramagnetic contribution to the
susceptibility (\ref{Fp}) takes place via intermediate intraband
transitions. Due to the smallness of the wavevectors $\left\vert \mathbf{q}%
_{1,2}\right\vert $\ compared with the characteristic lattice wavevectors $%
\left\vert \mathbf{q}_{1,2}\right\vert <<2\pi /a$\ the resonant interband
transitions in the field occur from a $-\mathcal{E}$\ negative energy level
to the positive $\mathcal{E}$\ energy level. The probabilities of intraband
transitions $\sim \partial n_{F}\left( s_{m},\mathbf{k}\right) /\partial 
\mathbf{k}$, and consequently, the intermediate intraband transitions take
place near the Fermi level. Thus, the nonlinear susceptibility tensor will
have maximal values if the involved frequencies are nearly resonant with the 
$2\varepsilon _{F}/\hbar $. This is the essence of the so-called Fermi-edge
resonance.

\subsection{Sum-frequency generation process in 2D hexagonal nanostructure}

Let us first consider the susceptibility of 2D hexagonal nanostructure $\chi
_{\alpha \beta \eta }\left( 2\omega ,2q_{x};\omega ,q_{x},\omega
,q_{x}\right) $ responsible for the second harmonic generation. Note that
the intensity of the second harmonic wave depends on the absolute value $%
|\chi _{\alpha \beta \eta }|$.\cite{Boyd2003} Due to intrinsic symmetry (\ref%
{symetri}) one can conclude that for second harmonic generation $\chi
_{\alpha \beta \eta }=\chi _{\alpha \eta \beta }$. First, let us compare our
result calculated for the FBZ with the DCA. In the DCA one can obtain
analytical results for zero temperature.\cite{Wang,Cheng} Thus, in the DCA
diamagnetic part is absent and in the leading order by $q$ from Eq. (\ref{Fp}%
) for the components of the second-order susceptibility tensor at $\omega
>>\gamma $ one can obtain%
\begin{equation*}
\chi _{xxx}(2\omega ;\omega ,\omega )=\frac{e^{3}\mathrm{v}_{F}^{2}q_{x}i}{%
\hbar ^{2}\omega ^{4}}
\end{equation*}%
\begin{equation}
\times \frac{3\varepsilon _{F}^{4}}{(\hbar ^{2}\omega ^{2}-4\varepsilon
_{F}^{2}+2\hbar ^{2}i\omega \gamma )(\hbar ^{2}\omega ^{2}-\varepsilon
_{F}^{2}+i\hbar ^{2}\omega \gamma )},  \label{Dxxx}
\end{equation}%
\begin{equation}
\chi _{xyy}(2\omega ;\omega ,\omega )=\chi _{xxx}(2\omega ;\omega ,\omega )%
\frac{\varepsilon _{F}^{2}-4\hbar ^{2}\omega ^{2}}{3\varepsilon _{F}^{2}},
\label{Dxyy}
\end{equation}%
\begin{equation}
\chi _{yyx}(2\omega ;\omega ,\omega )=\chi _{xxx}(2\omega ;\omega ,\omega )%
\frac{2\hbar ^{2}\omega ^{2}+\varepsilon _{F}^{2}}{3\varepsilon _{F}^{2}}.
\label{Dyyx}
\end{equation}%
These formulas\cite{kes} coincide with the results of [\onlinecite{Wang}] at 
$\gamma =0$ and the results of [\onlinecite{Cheng}] at $\omega >>\gamma $.
According to formulas (\ref{Dxxx}), (\ref{Dxyy}), and (\ref{Dyyx}) the
absolute values for all the components at resonance $\omega =\varepsilon
_{F}/\hbar $ coincide. In Fig. 3 and we plot nonzero components of the
second-order susceptibility tensor for silicene ($a=3.86\times 10^{-8}\ 
\mathrm{cm}$, $\gamma _{0}=1.087\ \mathrm{eV}$) and graphene ($a=2.46\times
10^{-8}\ \mathrm{cm}$, $\gamma _{0}=2.8\ \mathrm{eV}$). Note that $\chi
_{yxy}=\chi _{yyx}$. For comparison the results obtained from the DCA (\ref%
{Dxxx}), (\ref{Dxyy}), and (\ref{Dyyx}) are also shown. For both
nanostructures, we observe comparable values with main peaks near the Fermi
energy (second harmonic $\omega _{3}\simeq 2\varepsilon _{F}/\hbar $ ) and
double Fermi energy ($\omega _{1}=\omega _{2}\simeq 2\varepsilon _{F}/\hbar $%
). These are Fermi-edge resonances predicted in Refs. [%
\onlinecite{Wang,Cheng}]. For graphene the DCA is still valid since near the
Fermi energy $\varepsilon _{F}=700\ \mathrm{meV}$ isoenergy contours are
isotropic, and as a consequence, the FBZ integration is in agreement with
DCA. For silicene this Fermi energy is close to the nearest-neighbor hopping
energy $\varepsilon _{F}\thicksim \gamma _{0}$, the isoenergy contours are
nonisotropic and DCA is not valid which results in different maximal values.
To show the relative contributions of paramagnetic and diamagnetic
susceptibilities it is also plotted the ratio $\chi _{\alpha \beta \delta
}^{(p)}/\chi _{\alpha \beta \delta }^{(d)}$. As is seen near the Fermi-edge
resonances the main contribution is conditioned by the paramagnetic part.
Diamagnetic contribution to the susceptibility (\ref{Fd}) are conditioned by
the direct transitions $\sim n_{F}\left( 1,\mathbf{k}\right) -n_{F}\left( -1,%
\mathbf{k}\right) $\ without Fermi edge resonances. Since $\widehat{\tau }%
_{\alpha \beta }\sim \mathrm{v}_{F}^{2}/\gamma _{0}$,\ then for frequencies
smaller than nearest-neighbor hopping energy at the Fermi-edge resonance one
can conclude $\chi _{\alpha \beta \delta }^{(d)}/\chi _{\alpha \beta \delta
}^{(p)}\sim \hbar \gamma /\gamma _{0}<<1$\ and one can safely neglect the
diamagnetic contribution. For off-resonant high frequencies, the diamagnetic
part becomes comparable with the paramagnetic one. Note that in Refs. [%
\onlinecite{Wang, Cheng}] Fermi-edge resonances are attributed to the
resonant transitions in the linearly dispersed band structure of graphene.
As we see from Figs. 3 and 4, we have similar Fermi-edge resonances when the
DCA is no longer valid.

\begin{figure*}[tbp]
\includegraphics[width=.98\textwidth]{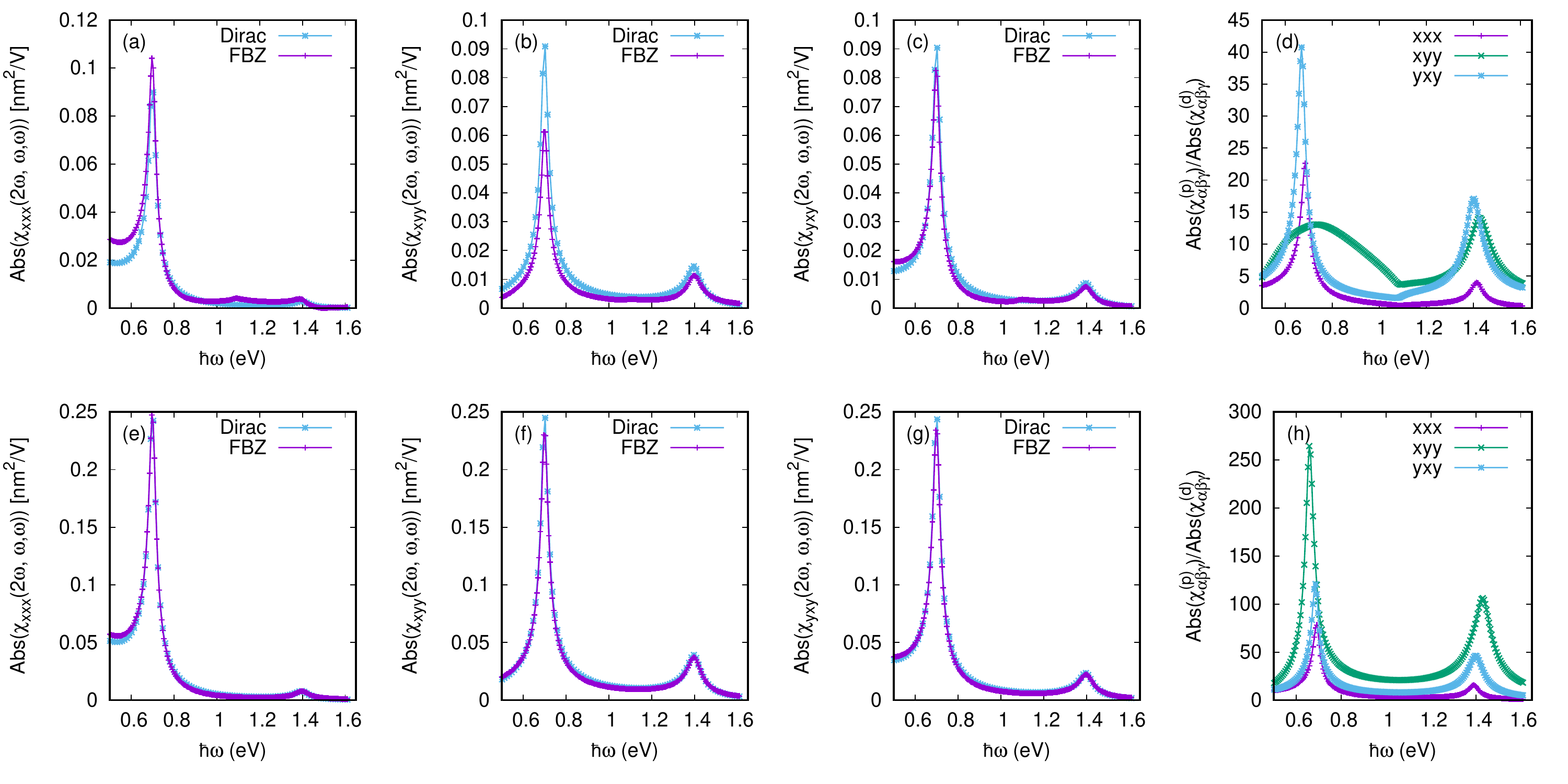}
\caption{The components of the second-order susceptibility tensor for the
process of second-harmonic generation as a function of the fundamental
frequency for silicene (a, b, c) and for graphene (e, f, g). The pump waves
are incident at $\protect\pi /4$. The Fermi energy is $\protect\varepsilon %
_{F}=700\ \mathrm{meV}$\textrm{.} The temperature is $k_{B}T=1\ \mathrm{meV}$%
. The relaxation rate is taken to be $\hbar \protect\gamma =30\ \mathrm{meV}$%
. The results obtained from the DCA are also shown. It is also plotted the
ratio $\protect\chi _{\protect\alpha \protect\beta \protect\delta }^{(p)}/%
\protect\chi _{\protect\alpha \protect\beta \protect\delta }^{(d)}$ for
silicene (d) and for graphene (h).}
\end{figure*}

\begin{figure*}[tbp]
\includegraphics[width=.98\textwidth]{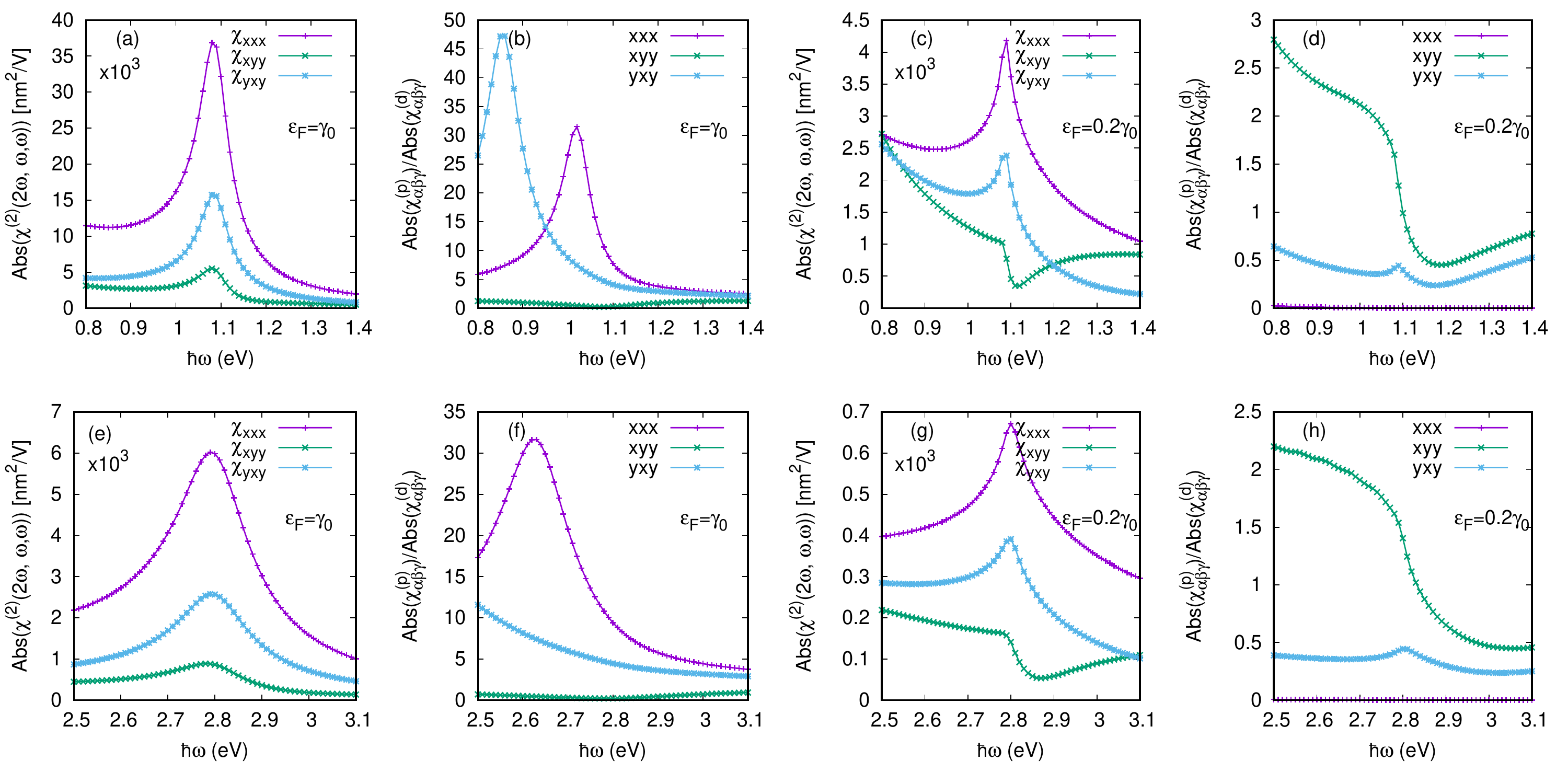}
\caption{The absolute value of components of the second-order susceptibility
tensor for the process of second-harmonic generation as a function of the
fundamental frequency near the van Hove singularity. The pump waves are
incident at $\protect\pi /4$. The temperature is $k_{B}T=3\ \mathrm{meV}$.
The relaxation rate is taken to be $\hbar \protect\gamma =0.05\protect%
\varepsilon _{F}$. (a, c) For silicene at Fermi energies $\protect%
\varepsilon _{F}=\protect\gamma _{0}$ and $\protect\varepsilon _{F}=0.2%
\protect\gamma _{0}$, respectively\textrm{.} (e, g) For graphene at $\protect%
\varepsilon _{F}=\protect\gamma _{0}$ and $\protect\varepsilon _{F}=0.2%
\protect\gamma _{0}$, respectively. It is also plotted the ratio $\protect%
\chi _{\protect\alpha \protect\beta \protect\delta }^{(p)}/\protect\chi _{%
\protect\alpha \protect\beta \protect\delta }^{(d)}$ for silicene (b,d) and
for graphene (f, h).}
\end{figure*}

For the higher frequencies, the other feature in the spectra which is
expected is the appearance of peaks due to van Hove singularity. The latter
takes place when one of the driving waves is in one photon resonance with
the van Hove singularity at the $M$ point of Brillouin zone. For this
propose in Fig. 4 we plot the components of the second-order susceptibility
tensor for silicene and graphene near the van Hove singularity for two Fermi
energies. For the first one, we rise the Fermi-edge resonance up to the
nearest-neighbor hopping energy: $\varepsilon _{F}=\gamma _{0}$. For the
second case, we take $\varepsilon _{F}=0.2\gamma _{0}$ to see pure van Hove
singularity. As is seen in both cases we have peaks. However, Fermi-edge
resonance peaks are larger by one order. This can be explained as follow. If
we consider cuts $S(\mathcal{E})$ of constant energy $\mathcal{E}$ in the
bandstructure, we can write $d^{2}\mathbf{k}=dSd\mathcal{E}/\left\vert
\partial \mathcal{E}/\partial \mathbf{k}\right\vert $. Near the $M$ point,
the density of states is high because of the van Hove singularity at the
saddle point ($\partial \mathcal{E}/\partial \mathbf{k}=0$), thereby one may
have an enhancement of the optical response of considered nanostructures. It
is well known that for the linear response the first order susceptibility%
\cite{Stauber} and for a third-harmonic generation, the nonlinear
susceptibility\cite{Hong} have resonant behavior for optical transitions
near the van Hove singularity irrespective to Fermi energy. In contrast to
the odd-order optical response, here due to inversion symmetry the peaks
near the van Hove singularity are not so pronounced. As expected from the
inversion symmetry of considered nanostructures at $\mathbf{q}_{1,2}=0$ we
have $\chi _{\alpha \beta \eta }\sim \int_{BZ}d\mathbf{k}\Phi _{\alpha \beta
\eta }\left( \mathbf{k}\right) =0$, where $\Phi _{\alpha \beta \eta }\left( 
\mathbf{k}\right) $ are the integrand functions in Eqs. (\ref{Fp}) and (\ref%
{Fd}). Therefore in the leading order by $\mathbf{q}$, we have $\chi
_{\alpha \beta \eta }\sim \mathbf{q}_{1,2}\int_{BZ}d\mathbf{k}\partial \Phi
_{\alpha \beta \eta }/\partial \mathbf{k}$. For paramagnetic part $\partial
\Phi _{\alpha \beta \eta }/\partial \mathbf{k}$ $\sim \partial
n_{F}/\partial \mathcal{E\cdot }\partial \mathcal{E}/\partial \mathbf{k}$
and the peaks at van Hove singularity are suppressed. The integrand
functions $\Phi _{\alpha \beta \eta }$ in Eq. (\ref{Fd}) also depend on the
transition matrix elements for velocity (\ref{velop}) and stress tensor (\ref%
{tauab}) operators which result in small peaks at van Hove singularity. In
Fig. 4 we also show the relative contributions of paramagnetic and
diamagnetic susceptibilities. As is seen near the van Hove singularity the
contributions of direct transitions in diamagnetic susceptibility are
significant.

In Fig 5, we plot results calculated with the parameters taken from the
experiment\cite{Zhang} by Zhang et al. For comparison with the experiment in
Fig. 5, we plotted the equivalent susceptibility for a bulk, which is
calculated dividing $\chi _{\alpha \beta \eta }$ by the effective thickness
of the monolayer. For both nanostructures we assume $d_{\text{eff}}\approx
0.3~\mathrm{nm}$. The results for graphene are in good agreement with
experiment\cite{Zhang} by Zhang et al. As is seen, for the fixed frequency
the susceptibility tensor grows rapidly as $\varepsilon _{F}$ approaches the
Fermi-edge resonances at one-photon ($2\varepsilon _{F}=\hbar \omega $) and
two-photon ($\varepsilon _{F}=\hbar \omega $) energies.

\begin{figure}[tbp]
\includegraphics[width=.4\textwidth]{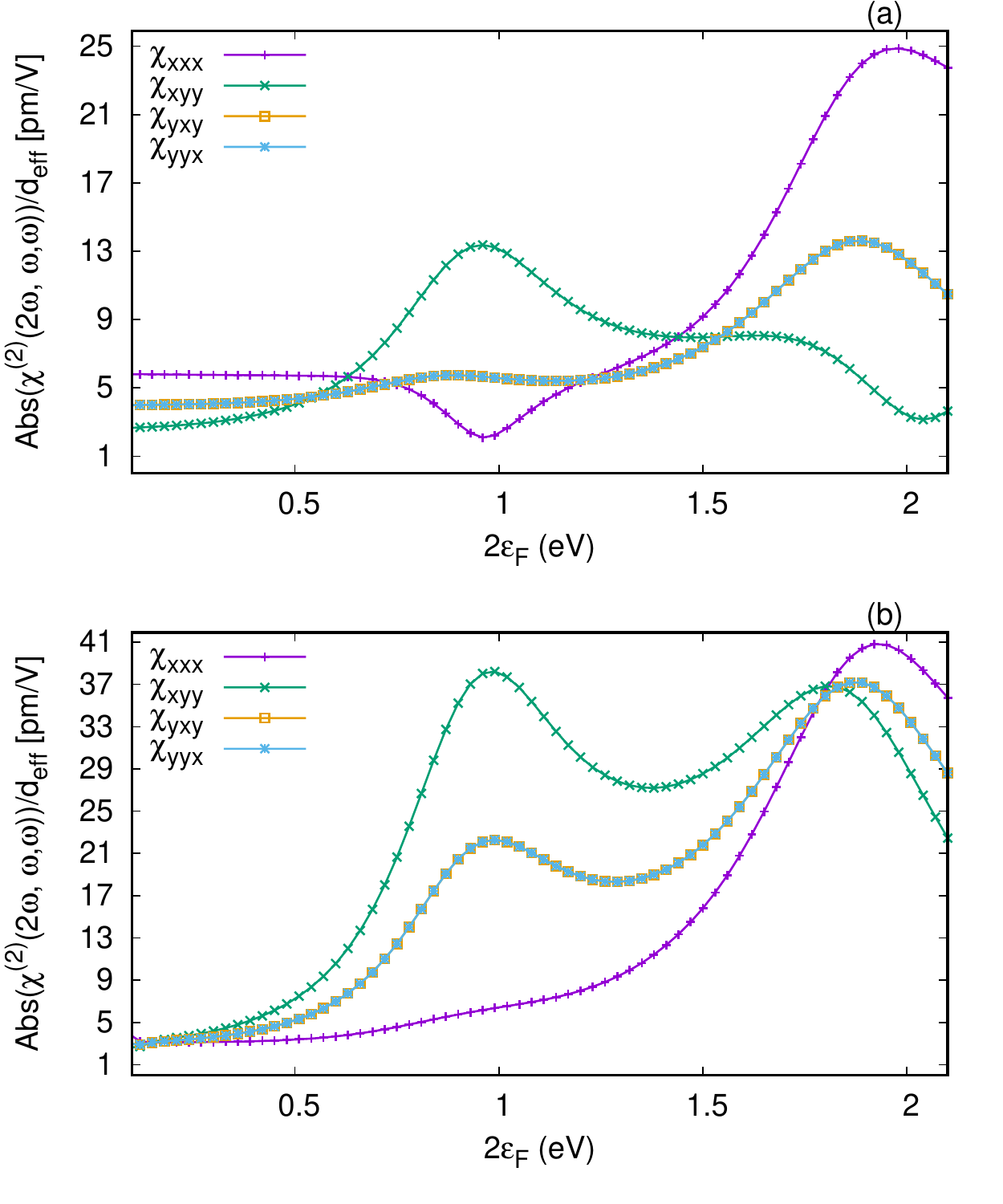}
\caption{The second-order susceptibility tensor components for the process
of second-harmonic generation as a function of the Fermi energy for silicene
(a) and for graphene (b). The frequency is $\hbar \protect\omega =0.95\ 
\mathrm{eV}$. The pump waves are incident at $\protect\pi /3$. The
temperature is $k_{B}T=26\ \mathrm{meV}$. The relaxation rate is taken to be 
$\hbar \protect\gamma =0.2\protect\varepsilon _{F}$.}
\end{figure}

In Fig. 6, the maximum values of the second-order susceptibility tensor
components for the process of second harmonic generation as a function of
the Fermi energy for silicene and graphene are shown. As is seen, with the
increase of Fermi energy and consequently resonant frequency the maximum
value of susceptibility tensor is reduced. For graphene one can interpolate
the dependence $\chi _{\alpha \beta \eta }\thicksim 1/\varepsilon _{F}^{2}$,
which is also clear from the analytical results (\ref{Dxxx}), (\ref{Dxyy}),
and (\ref{Dyyx}). For silicene, the interpolation $\chi _{\alpha \beta \eta
}\thicksim 1/\varepsilon _{F}^{2}$ is valid up to energies $\varepsilon
_{F}\simeq 0.7\gamma _{0}$. From the inset of Fig. 6(a) we see that for
silicene near $1\ \mathrm{eV}$\ we have a local maximum for $\chi _{xxx}$.
This behavior reflects the van Hove singularity at $\varepsilon _{F}\simeq
\gamma _{0}$.

\begin{figure}[tbp]
\includegraphics[width=.4\textwidth]{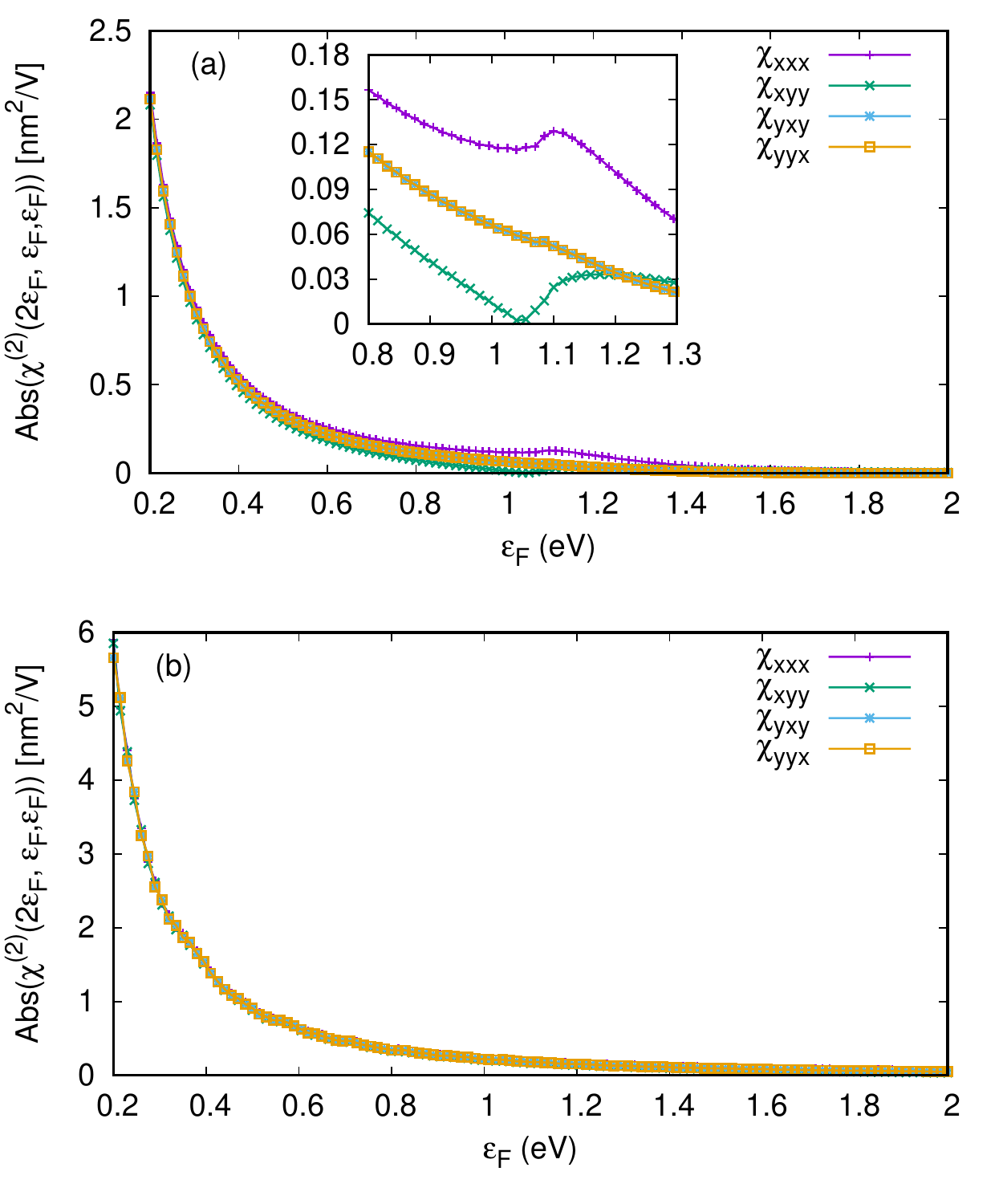}
\caption{The maximum values of the second-order susceptibility tensor
components for the process of second-harmonic generation at the resonant
pump frequency $\protect\omega _{1}=\protect\omega _{2}=\protect\varepsilon %
_{F}/\hbar $ as a function of the Fermi energy for silicene (a) and for
graphene (b). The pump waves are incident at $\protect\pi /3$. The
temperature is $k_{B}T=3\ \mathrm{meV}$. The relaxation rate is taken to be $%
\hbar \protect\gamma =5\ \mathrm{meV}$.}
\end{figure}

In general, to clear up the deviations of the second-order susceptibility
tensor calculated for the FBZ from the DCA, in Fig. 7 we plot the ratios of
the absolute values of susceptibility tensor components for the process of
the second-harmonic generation calculated with and without DCA (for nonzero
temperature) as a function of the Fermi energy scaled to the
nearest-neighbor hopping energy $\gamma _{0}$. The relaxation rate and
temperature are also scaled. Although Eq. (\ref{CI2}) depends on the lattice
spacing $a$, however since the general $\chi ^{(2)}$\ and susceptibility
tensor $\chi _{D}^{(2)}$, calculated in the DCA $\sim q/k_{b}$, the ratios
are independent of lattice spacing $a$, and Fig. 7 is applicable to all
considered hexagonal nanostructures if $\left\vert \mathbf{q}%
_{1,2}\right\vert <<2\pi /a$. As we see, the DCA is valid up to Fermi
energies $\varepsilon _{F}=0.4\gamma _{0}$\ and there are considerable
qualitative and quantitative deviations close to the van Hove singularity.%
\textrm{\ }Near the van Hove singularity, the nonlinear susceptibility
tensor $\chi _{\alpha \beta \eta }$\ for the second harmonic generation
process strongly depends on the product of three velocity matrix elements (%
\ref{vnm})%
\begin{equation*}
\Pi _{\alpha \beta \eta }\left( \mathbf{k}\right) =\langle -1,\mathbf{k}|%
\widehat{\mathrm{v}}_{\alpha }|1,\mathbf{k}\rangle \langle 1,\mathbf{k}|%
\widehat{\mathrm{v}}_{\beta }|1,\mathbf{k}\rangle \langle -1,\mathbf{k}|%
\widehat{\mathrm{v}}_{\eta }|-1,\mathbf{k}\rangle
\end{equation*}%
\textrm{\ }along the critical energy isoline $\mathcal{E}=\gamma _{0}$. This
product describes the main transition (see Fig. 2) when $\omega _{1}=\omega
_{2}\equiv \omega $\ and $\omega _{3}=2\omega $. The $\Pi _{xyy}\left( 
\mathbf{k}\right) $\ vanishes along the critical energy isoline and,
therefore $\chi _{xyy}\simeq 0$\ near the van Hove singularity. On the
contrary, $\Pi _{xxx}\left( \mathbf{k}\right) $\ is nonvanishing, and
therefore $\chi _{xxx}\ $has a peak.

We have also investigated the temperature dependence of the maximum values
of the second-order susceptibility tensor components for second harmonic
generation process. The latter is plotted for graphene in Fig. 8 at the
various Fermi energies. The same qualitative picture we have for the
silicene. From Fig. 8 one can interpolate the dependence $\chi _{\alpha
\beta \eta }\thicksim 1/T$. The latter strictly restricts the second
harmonic generation process at the room temperatures. The maximum value of
the calculated second-order susceptibility for the second harmonic
generation at the resonances are: $|\chi _{\alpha \beta \eta }\left( 2\omega
;\omega ,\omega \right) |_{\max }\simeq 0.1-0.4$ $\mathrm{nm}^{2}\mathrm{/V}$%
. Thus, for the bulk we obtain $|\chi _{\alpha \beta \eta }\left( 2\omega
;\omega ,\omega \right) |_{\max }/d_{\text{eff}}\simeq 300-1000$ $\mathrm{%
pm/V}$. Compared with common materials for the second-order nonlinearity
these values are very large. For\textrm{\ }the lithium niobate $\chi
^{(2)}\simeq 20$ $\mathrm{pm/V.}$\textrm{\ }

\begin{figure}[tbp]
\includegraphics[width=.4\textwidth]{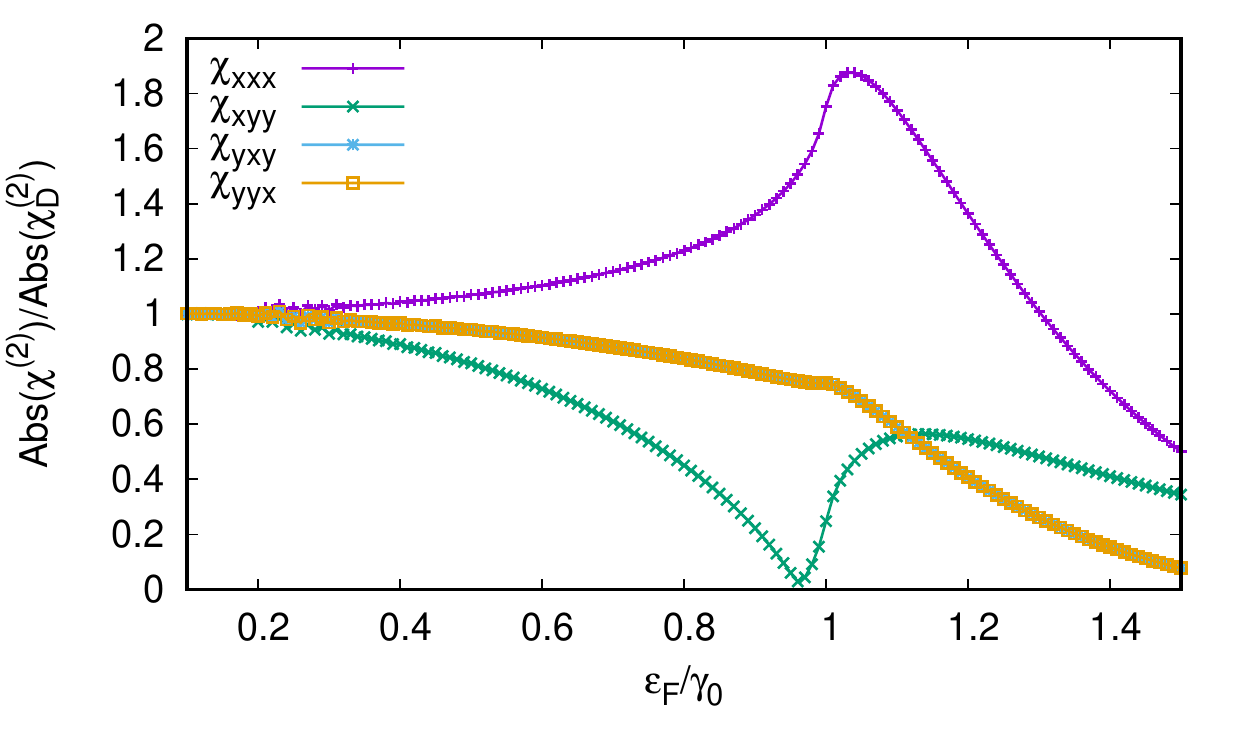}
\caption{The ratios of the absolute values of susceptibility tensor
components for the process of second-harmonic generation calculated with and
without DCA as a function of the Fermi energy scaled to nearest-neighbor
hopping energy $\protect\gamma _{0}$. The pump waves are incident at $%
\protect\pi /3$ with resonant frequency $\hbar \protect\omega =\protect%
\varepsilon _{F}$. The relaxation rate and the temperature are taken to be $%
\hbar \protect\gamma /\protect\varepsilon _{F}=0.01$ and $k_{B}T/\protect%
\varepsilon _{F}=0.003$, respectively.}
\end{figure}

\begin{figure}[tbp]
\includegraphics[width=.4\textwidth]{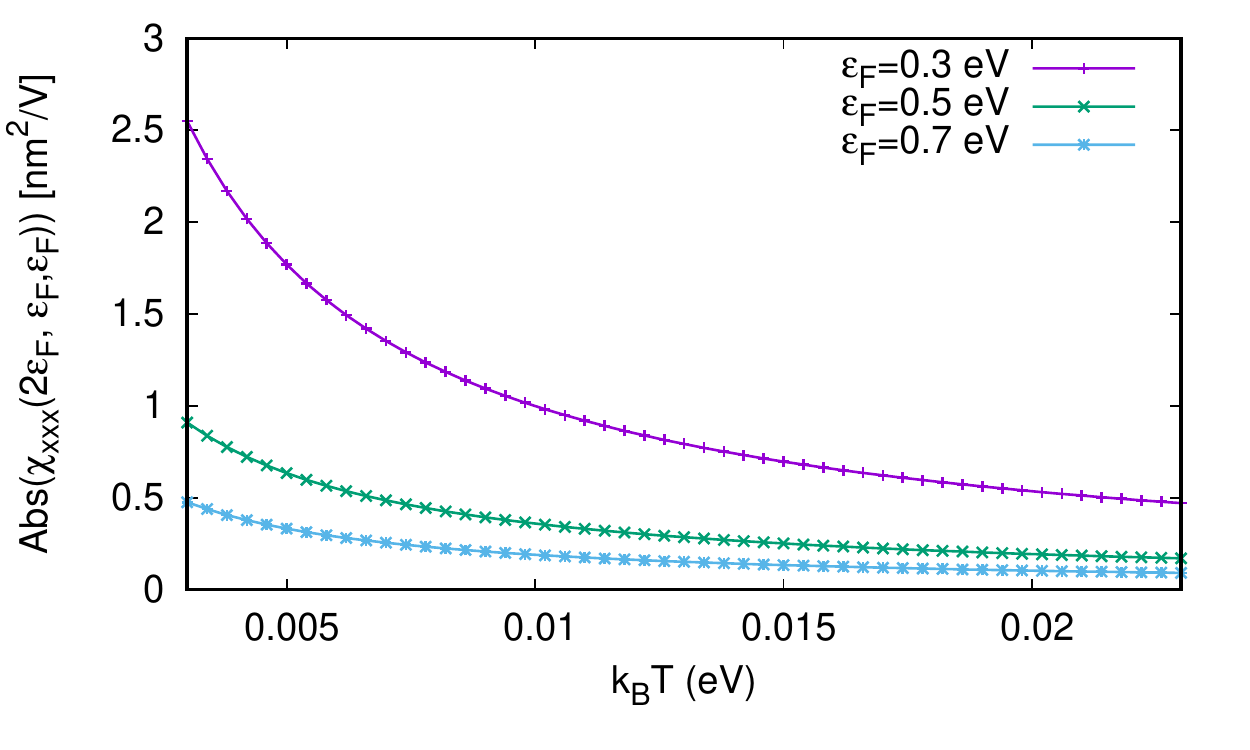}
\caption{The maximum values of the second order susceptibility tensor
components for the process of second-harmonic generation as a function of
the temperature for graphene at various Fermi energies. The pump waves are
incident at $\protect\pi /3$. The relaxation rate is taken to be $\hbar 
\protect\gamma =5\ \mathrm{meV}$.}
\end{figure}

\begin{figure}[tbp]
\includegraphics[width=.4\textwidth]{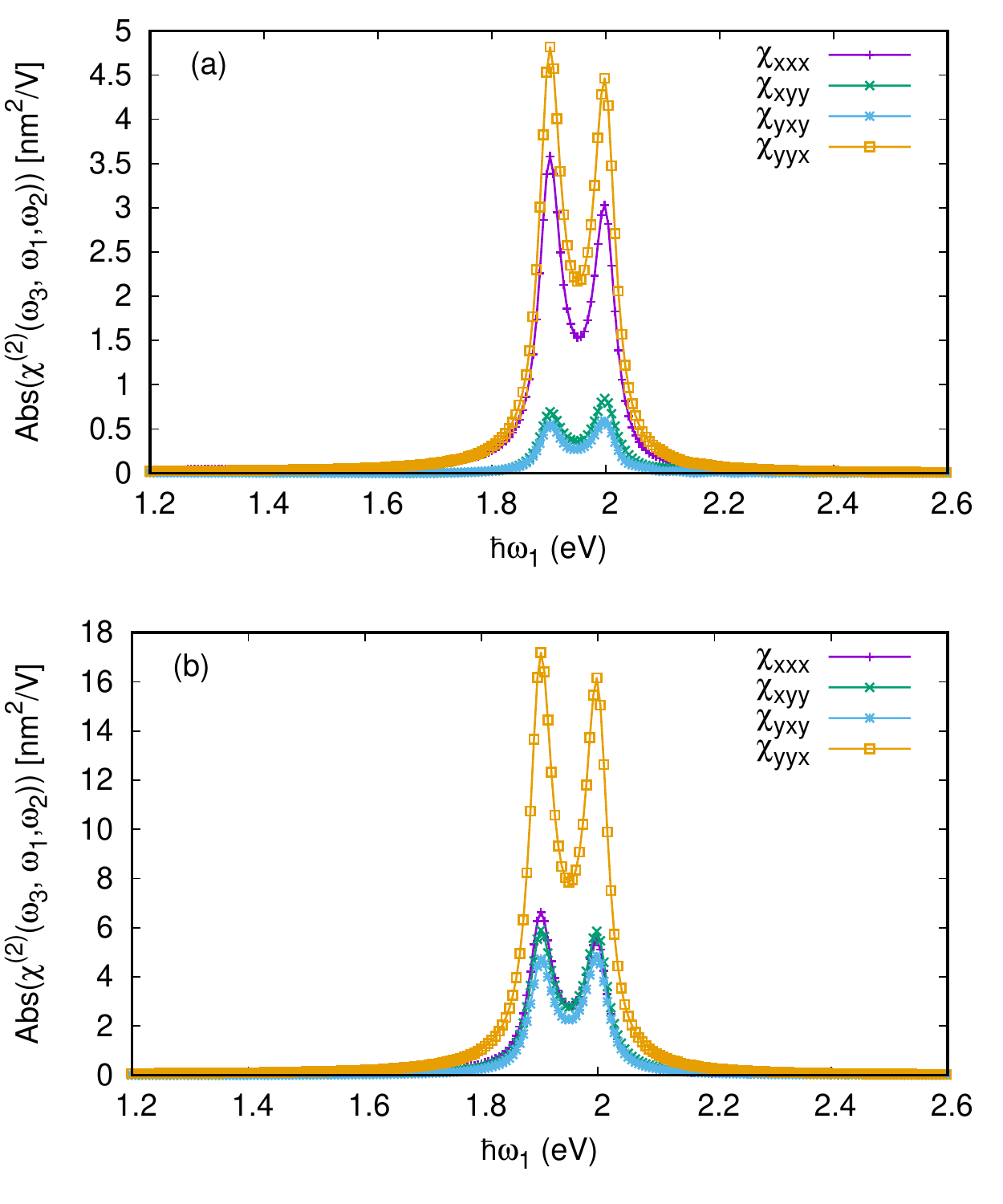}
\caption{The absolute values of susceptibility tensor components responsible
for sum-frequency generation as a function of one of the pump frequencies $%
\protect\omega _{1}$ for silicene (a) and for graphene (b). The frequency $%
\protect\omega _{2}$ is fixed at $0.1\ \mathrm{eV}/\hbar $. The pump waves
are incident at $\protect\pi /3$. The Fermi energy is $\protect\varepsilon %
_{F}=1\ \mathrm{eV}$\textrm{.} The temperature is $k_{B}T=3\ \mathrm{meV}$.
The relaxation rate is taken to be $\hbar \protect\gamma =5\ \mathrm{meV}$.}
\end{figure}

For the second harmonic generation, we have resonance when the output
radiation is close to $2\varepsilon _{F}/\hbar $. In case, when one of the
pump frequencies is very small compared to other: $\omega _{2}<<\omega _{1}$%
, one can realize a double resonance $\omega _{1}\simeq \omega _{3}\thicksim
2\varepsilon _{F}/\hbar $ with the considerable enhancement of the output
nonlinear response. Thus, in Fig. 9 the absolute values of susceptibility
tensor components responsible for sum-frequency generation as a function of
one of the pump frequencies $\omega _{1}$ for silicene and for graphene are
displayed. As is seen, we have maximal enhancement when the low-frequency
pump wave is p-polarized. At that, the equivalent susceptibility for bulk is 
$|\chi _{\alpha \beta \eta }|_{\max }/d_{\text{eff}}\simeq 1.5\times
10^{4}-6\times 10^{4}$ $\mathrm{pm/V}$. Thus, at the double resonance
susceptibility reaches huge values that is more pronounced for difference
frequency generation process. In this case, the contribution of the
diamagnetic part (\ref{Fd}) is much less due to double resonance.

\subsection{Difference frequency generation processes: generation of plasmons%
}

It is also of interest the difference frequency generation processes in the
considered nanostructures, since it can be used for all-optical generation
of plasmons or THz radiation from visible light. For this propose we examine
the susceptibility tensor $\chi _{\alpha \beta \eta }\left( \omega _{3},%
\mathbf{q}_{3};\omega _{1},\mathbf{q}_{1},-\omega _{2},-\mathbf{q}%
_{2}\right) $. In Fig. 10 the absolute values of susceptibility tensor
components responsible for difference frequency generation as a function of
the pump frequency $\omega _{1}$ at the fixed idler frequency $\hbar \omega
_{3}=0.1\varepsilon _{F}$ are plotted for graphene and silicene. The maximal
values of susceptibility tensors correspond to the cases when the output
radiation is p-polarized. As is seen from this figure, even for such high
frequency pump and signal waves the both nanostructures exhibit large values
of $|\chi _{\alpha \beta \eta }|_{\max }/d_{\text{eff}}\simeq 1.5\times
10^{4}$ $\mathrm{pm/V}$\textrm{. } 
\begin{figure}[tbp]
\includegraphics[width=.4\textwidth]{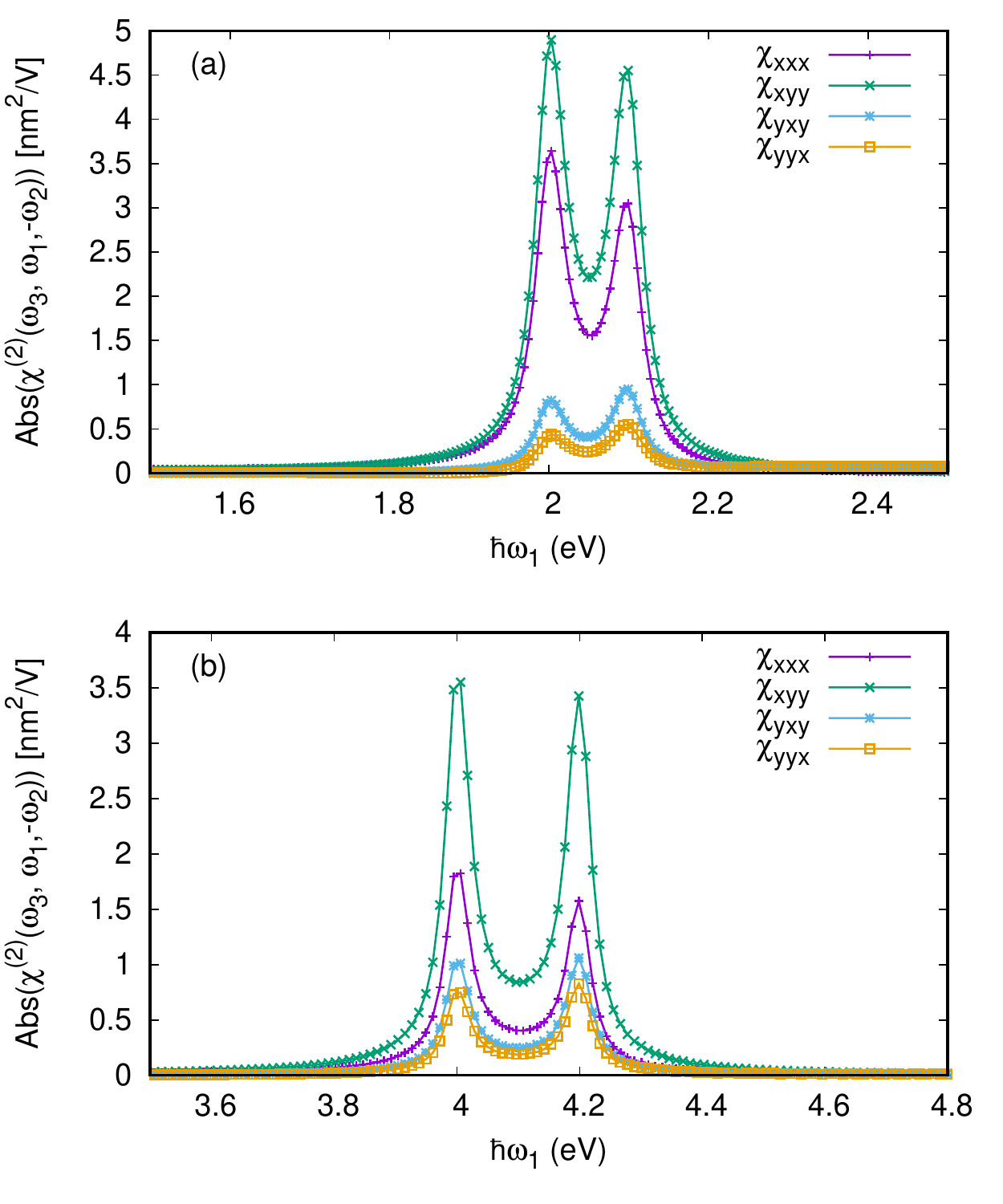}
\caption{The absolute values of susceptibility tensor components responsible
for difference frequency generation as a function of the pump frequency $%
\protect\omega _{1}$ at the fixed idler frequency $\protect\omega _{3}=0.1%
\protect\varepsilon _{F}/\hbar $. The pump and the signal waves are incident
at $\protect\pi /3$. The temperature is $k_{B}T=3\ \mathrm{meV}$. The
relaxation rate is taken to be $\hbar \protect\gamma =5\ \mathrm{meV}$. (a)
For silicene at $\protect\varepsilon _{F}=1\ \mathrm{eV}$ and (b) for
graphene at $\protect\varepsilon _{F}=2\ \mathrm{eV}$\textrm{.}}
\end{figure}
\begin{figure}[tbp]
\includegraphics[width=.4\textwidth]{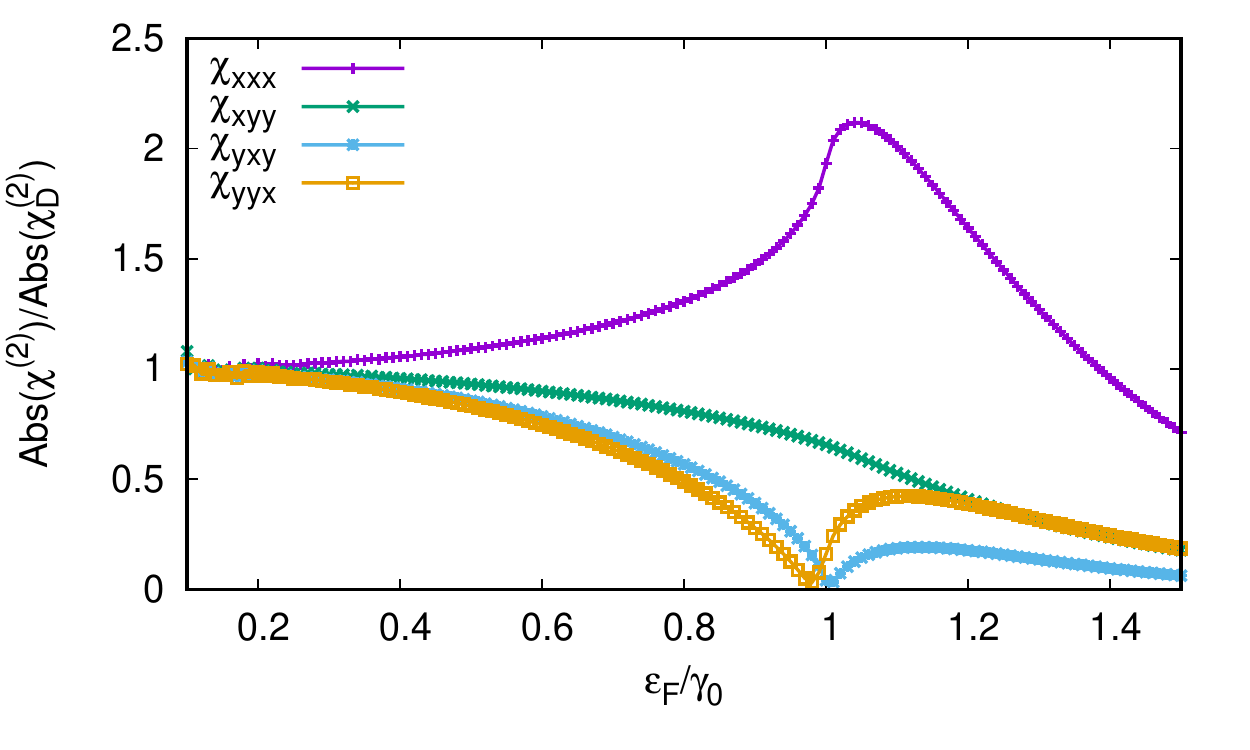}
\caption{The ratios of the absolute values of susceptibility tensor
components for difference frequency generation calculated with and without
DCA as a function of the Fermi energy scaled to nearest-neighbor hopping
energy $\protect\gamma _{0}$. The pump and the signal waves are incident at $%
\protect\pi /3$ with frequencies $\hbar \protect\omega _{1}=2\protect%
\varepsilon _{F}$ and $\hbar \protect\omega _{2}=\hbar \protect\omega %
_{1}-0.1\protect\varepsilon _{F}$. The relaxation rate and the temperature
are taken to be $\hbar \protect\gamma /\protect\varepsilon _{F}=0.01$ and $%
k_{B}T/\protect\varepsilon _{F}=0.003$, respectively.}
\end{figure}

In Fig. 11, we plot the ratios of the absolute values of susceptibility
tensor components for difference frequency generation calculated with and
without DCA as a function of the Fermi energy scaled to nearest-neighbor
hopping energy $\gamma _{0}$. The pump and the signal waves are taken\ with
frequencies $\hbar \omega _{1}=2\varepsilon _{F}$\ and $\hbar \omega
_{2}=\hbar \omega _{1}-0.1\varepsilon _{F}$. As we see, the DCA is valid up
to Fermi energies $\varepsilon _{F}=0.4\gamma _{0}$\ and there is a
considerable deviation when one approaches the van Hove singularity. In this
case the nonlinear susceptibility tensor $\chi _{\alpha \beta \eta }$\ for
the difference frequency generation process strongly depends on the product
of three velocity matrix elements%
\begin{equation*}
\Pi _{\alpha \beta \eta }\left( \mathbf{k}\right) =\langle 1,\mathbf{k}|%
\widehat{\mathrm{v}}_{\alpha }|1,\mathbf{k}\rangle \langle -1,\mathbf{k}|%
\widehat{\mathrm{v}}_{\beta }|1,\mathbf{k}\rangle \langle 1,\mathbf{k}|%
\widehat{\mathrm{v}}_{\eta }|-1,\mathbf{k}\rangle
\end{equation*}%
along the energy isoline $\mathcal{E}=\varepsilon _{F}$. This product
describes the main transition (see Fig. 2) when $\omega _{1}\simeq \omega
_{2}\simeq 2\varepsilon _{F}$\ and $\omega _{3}<<\omega _{1}$. Along the
critical energy isoline ($\mathcal{E}=\gamma _{0}$) $\left\vert \Pi
_{yxy}\left( \mathbf{k}\right) \right\vert \simeq \left\vert \Pi
_{yyx}\left( \mathbf{k}\right) \right\vert \simeq 0$\ and, therefore $\chi
_{yxy}\simeq \chi _{yyx}\simeq 0$\ near the van Hove singularity. On the
contrary, $\Pi _{xxx}\left( \mathbf{k}\right) $\ is nonvanishing, and
therefore $\chi _{xxx}\ $has a peak. 
\begin{figure}[tbp]
\includegraphics[width=.4\textwidth]{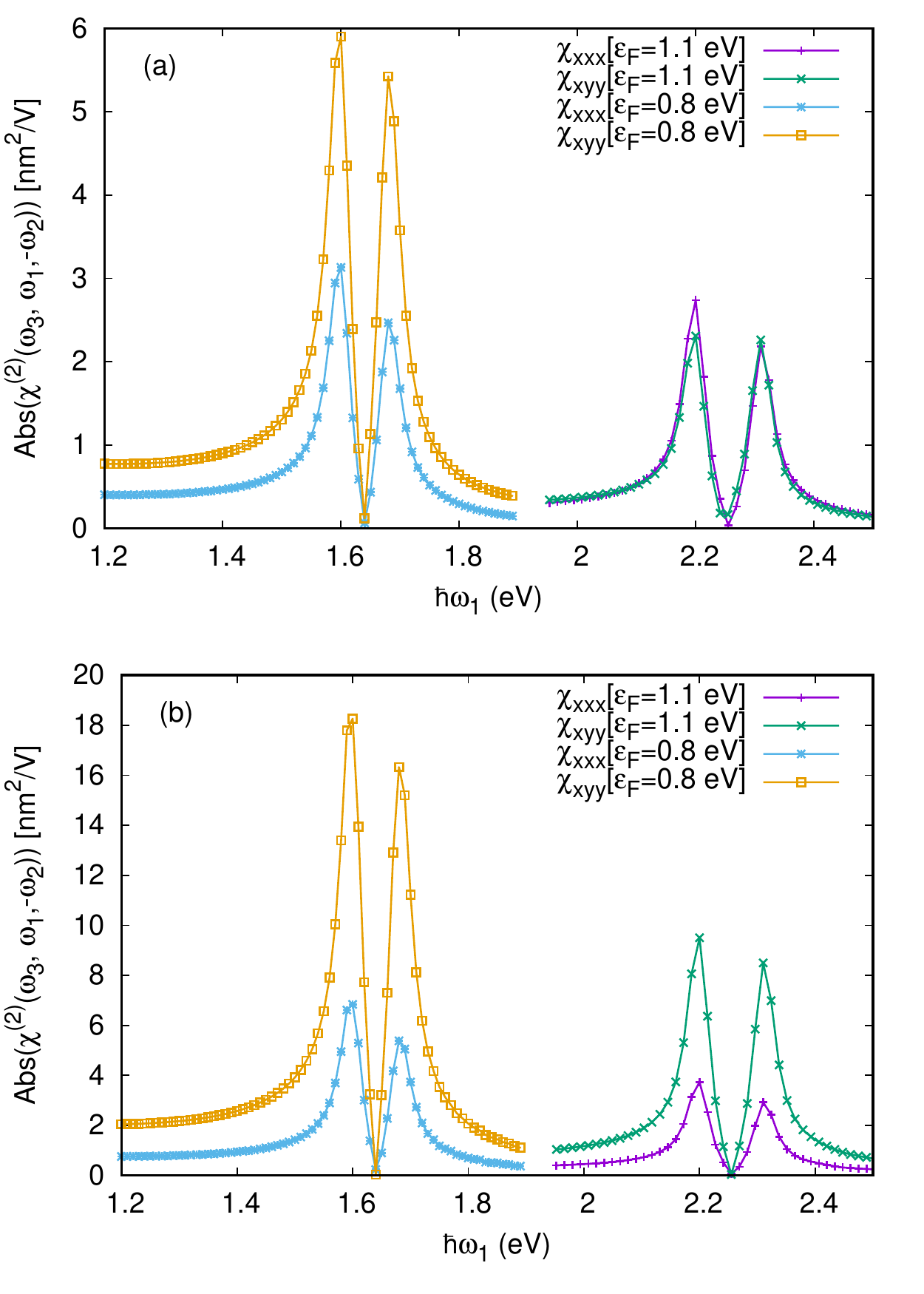}
\caption{The absolute values of susceptibility tensor components responsible
for plasmon generation as a function of the pump frequency $\protect\omega %
_{1}$ at the fixed idler frequency $\protect\omega _{3}=\protect\omega %
_{p}=0.1\protect\varepsilon _{F}/\hbar $ for various Fermi energies. In
plane wave vectors of the pump and the signal waves are opposite. The
temperature is $k_{B}T=2\ \mathrm{meV}$. The relaxation rate is taken to be $%
\hbar \protect\gamma =5\ \mathrm{meV}$. (a) For silicene and (b) for graphene%
\textrm{.}}
\end{figure}
\begin{figure}[tbp]
\includegraphics[width=.5\textwidth]{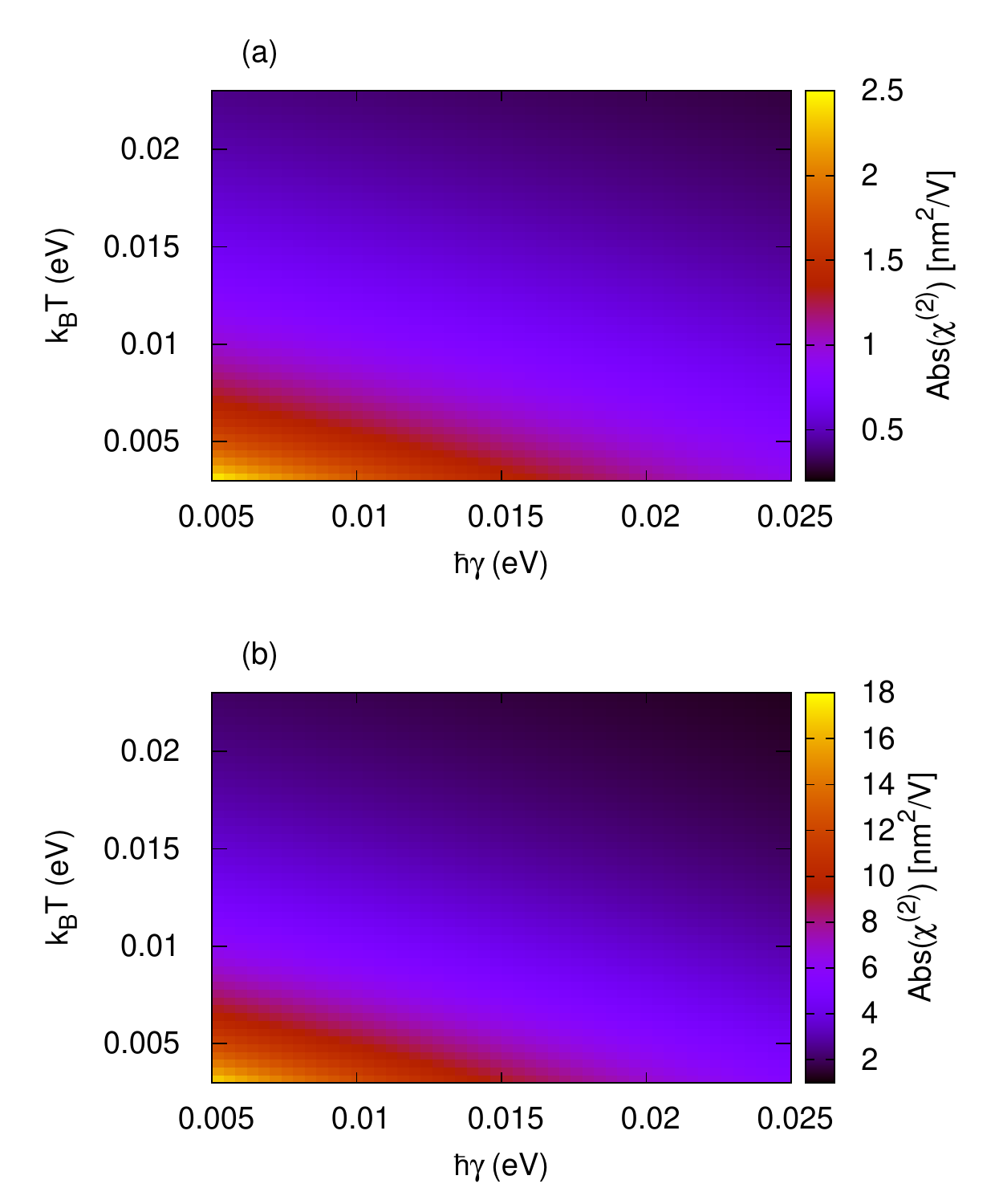}
\caption{The density plot of the maximum values of susceptibility tensor
component $\protect\chi _{xxx}$ responsible for plasmon generation as a
function of the temperature and relaxation rate. The pump frequency $\protect%
\omega _{1}=2\protect\varepsilon _{F}/\hbar $ and the signal frequency is $%
\protect\omega _{2}=1.9\protect\varepsilon _{F}/\hbar .$(a) For silicene at
Fermi energy $0.8\ \mathrm{eV}$ and (b) for graphene\textrm{\ }at Fermi
energy $0.5\ \mathrm{eV.}$}
\end{figure}

Next, we consider double resonant plasmon generation with the oblique
incidence of pump and signal electromagnetic waves. For graphene, the
effective spin-orbit coupling is negligibly small. However for silicene,
germanene, and stanene, spin-orbit coupling opens gap $\varepsilon _{soc}$.
For silicene $\varepsilon _{soc}\simeq 8\ \mathrm{meV.}$ At $\varepsilon
_{F}>>\varepsilon _{soc}$ in case of graphene and silicene we can use the
following dispersion relation for plasmon: 
\begin{equation}
\hbar \omega _{p}\left( q\right) =\sqrt{\frac{2\alpha \varepsilon _{F}\hbar
cq}{\epsilon }},  \label{plasmon}
\end{equation}%
where $q$ is the wave vector, $\alpha =1/137$ is the fine structure
constant. Here, $\epsilon \equiv \left( \epsilon _{1}+\epsilon _{2}\right)
/2 $, with the dielectric constants of the above $\epsilon _{1}$ and below $%
\epsilon _{2}$ surrounding media. For the plasmon generation we need to
satisfy the phase-matching conditions:%
\begin{equation*}
\hbar \omega _{1}-\hbar \omega _{2}=\hbar \omega _{p}\left( q\right) ,
\end{equation*}%
and 
\begin{equation*}
\mathbf{q}_{1}-\mathbf{q}_{2}=\mathbf{q.}
\end{equation*}%
Assuming that pump and signal waves are incident from the vacuum ($\epsilon
_{1}=1$) and all in-plane photon wave vectors are directed along the $x$%
-axis with $\vartheta _{2}=\pi -\vartheta _{1}$, for the resonant incident
angle we will have%
\begin{equation*}
\cos \vartheta _{1}=\frac{\epsilon }{2\alpha \varepsilon _{F}}\frac{\hbar
^{2}\omega _{p}^{2}}{\hbar \omega _{1}+\hbar \omega _{2}}.
\end{equation*}%
We will assume a silicon dioxide substrate ($\epsilon =2.75$). In Fig. 12
the absolute values of susceptibility tensor components responsible for
plasmon generation as a function of the pump frequency $\omega _{1}$ at the
fixed idler frequency $\hbar \omega _{3}=\hbar \omega _{p}=0.1\varepsilon
_{F}$ for various Fermi energies are displayed. Near the resonant
frequencies $\hbar \omega _{1}\simeq \hbar \omega _{2}\simeq 2\varepsilon
_{F}$, the resonant incident angle is $\vartheta _{1}\simeq \pi /3$. As is
seen from Fig. 12, the plasmon generation is more preferable by the
s-polarized waves. For both nanostructures the maximum value of the
calculated second-order susceptibility for the plasmon generation processes
due to the double resonance can reach huge values as high as $|\chi _{\alpha
\beta \eta }|_{\max }\simeq 20$ $\mathrm{nm}^{2}\mathrm{/V}$\textrm{. }

We have also investigated the temperature and the relaxation rate dependence
on the maximum values of the second-order susceptibility tensor components
for the plasmon generation process. Figure 13 represents the density plot of
the maximum values of susceptibility tensor component $\chi _{xxx}$
responsible for plasmon generation as a function of the temperature and
relaxation rate. The pump frequency $\hbar \omega _{1}=2\varepsilon _{F}$
and the signal frequency is $\hbar \omega _{2}=1.9\varepsilon _{F}$. From
Fig. 13 one can interpolate the dependence $\chi _{\alpha \beta \eta
}\thicksim 1/\left( T^{6/5}\gamma ^{1/2}\right) $ .

Let us make some estimation and compare our results with the other ones. The
maximum value of the calculated second-order susceptibility for the plasmon
generation processes corresponds to a bulk of $\sim 10^{5}$ $\mathrm{pm/V}$%
\textrm{. }In this case the\textrm{\ }off-resonance susceptibility $|\chi
_{\alpha \beta \eta }|_{\mathrm{off}}\simeq $ $3$ $\mathrm{nm}^{2}\mathrm{/V}
$, which corresponds to a bulk of $\sim 10^{4}$ $\mathrm{pm/V}$. Regarding
the experimental results. Constant et al.\cite{Constant} reported a bulk
susceptibility $10^{5}$ $\mathrm{pm/V}$ for off resonant plasmon generation
with the waves of frequencies $\simeq 2$ $\mathrm{eV}$ and doping level $%
\varepsilon _{F}=0.5$ $\mathrm{eV}$. The reported value is close to our
theoretical result but for resonant susceptibility. Our off resonant
susceptibility is order of magnitude smaller than the experimental one. As
was mentioned above, at off-resonant generation of plasmons one should take
into account the many-body effects. In particular, multiple hot carrier
generation.\cite{hot} In this case the independent carrier picture is not
applicable.

Regarding the other hexagonal nanostructures. Germanene and stanene have
comparable to silicene Fermi velocities and nearest-neighbor hopping
energies $\sim 1$ $\mathrm{eV}$\ and the results for these materials will be
close to those of the silicene. However one should take into account that
for germanene and stanene the spin-orbit coupling opens a gap $\sim 0.1$ $%
\mathrm{eV}$, and the results obtained in the scope of our approach can be
applicable at the sufficiently high doping and involved frequencies $>0.2$ $%
\mathrm{eV}$.

\section{Conclusion}

We have developed a microscopic quantum ansatz for analytical and numerical
calculation of the second-order nonlinear response of hexagonal 2D
nanostructures (graphene and its analogs -silicene, germanene, and stanene)
beyond the Dirac cone approximation, which is applicable to the excitations
in the full Brillouin zone. The second-order nonlinear optical
susceptibility tensor has been calculated for monolayers of graphene and
silicene. We have taken into account triangular (paramagnetic part) and
nonlinear bubble diagrams (diamagnetic part) for second-order nonlinear
optical susceptibility. The latter is absent in the Dirac cone
approximation. We have demonstrated that Fermi-edge resonances also take
place for the high-frequency excitations beyond the linear dispersion of
massless Dirac fermions and are conditioned by the paramagnetic part of
nonlinear optical susceptibility. For off-resonant high frequencies, the
diamagnetic part becomes comparable with the paramagnetic one. The Dirac
cone approximation is valid up to the Fermi energies $\varepsilon
_{F}=0.4\gamma _{0}$\ and there are considerable qualitative and
quantitative deviations when one approaches the van Hove singularity.\textrm{%
\ }The van Hove singularity is not so pronounced as in the case of odd-order
optical responses. For visible and UV frequencies both nanostructures
exhibit a large second-order response. For the difference/sum-frequency
generation processes, one can realize double resonance -- when the pump wave
frequency and the idler frequency are close to double Fermi energy-- the
second-order susceptibility reaches huge values. The obtained results show
that along with graphene at sufficiently high doping silicene, germanene,
and stanene are promising materials for optoelectronic applications. In
particular, these materials are ideally suited for the all-optical plasmon
generation at the double Fermi-edge resonances. We have also investigated
temperature ($T$) and relaxation rate ($\gamma $) dependences of the
second-order susceptibility tensor components for the process of plasmon
generation which in the wide range show the dependence $\chi _{\alpha \beta
\eta }\thicksim 1/\left( T^{6/5}\gamma ^{1/2}\right) $.

\begin{acknowledgments}
This work was supported by the RA State Committee of Science and Belarusian
Republican Foundation for Fundamental Research (RB) in the frame of the
joint research project SCS 18BL-020.
\end{acknowledgments}

\end{document}